\documentclass[aps,twocolumn,pra,superscriptaddress,nofootinbib]{revtex4-2}
\pdfoutput=1

\usepackage{booktabs}
\usepackage{amsmath}
\usepackage{amssymb}
\usepackage{amsthm}
\usepackage{bm, dsfont}
\usepackage[usenames,dvipsnames]{color}
\usepackage{colortbl}
\usepackage[table]{xcolor}
\usepackage{enumitem}

\usepackage{graphicx}
\usepackage{natbib}
\usepackage[colorlinks=true,linkcolor=blue,citecolor=blue,urlcolor=blue]{hyperref}
\usepackage{cleveref}
\usepackage{hypcap}
\usepackage{verbatim, float}
\usepackage[normalem]{ulem}
\usepackage{physics}		

\newcommand{\id}{\mathds{1}}
\newcommand{\ii}{\mathrm{i}}

\newcommand{\be}{\begin{equation}}
\newcommand{\ee}{\end{equation}}

\newcommand{\cN}{\mathcal{N}}
\newcommand{\mean}[1]{\mathds{E}\left[#1\right]}

\newcommand{\CP}{\bigtimes}
\usepackage{mathtools}

\def\P{{\mathbb{P}}}
\def\E{{\mathbb{E}}}

\newtheorem{theorem}{Theorem}
\newtheorem*{definition}{Definition}
\newtheorem{corollary}{Corollary}[theorem]

\newtheorem{proposition}[theorem]{Proposition}
\newtheorem{remark}{Remark}[theorem]

\begin{document}



\title{Bell nonlocality in quantum networks with unreliable sources:
\\  Loophole-free postelection via self-testing}

\author{Sadra Boreiri}
\author{Nicolas Brunner}
\author{Pavel Sekatski}
\affiliation{Department of Applied Physics, University of Geneva, Switzerland}

 \begin{abstract}

We discuss Bell nonlocality in quantum networks
with unreliable sources. Our main result is a condition on the observed data which ensures that inconclusive events can be safely discarded, without introducing any loophole. More formally, we characterize the fair-sampling property for measurements in a network. When all measurements are fair-sampling, we show that the post-selection of conclusive outcomes does not compromise the assumption of source independence, hence avoiding the detection loophole. Furthermore, we show that in some cases, the fair-sampling property can in fact be guaranteed based only on observed data. To show this, we prove that saturation of the Finner inequality provides a self-test of the underlying quantum model. We illustrate the relevance of our results by demonstrating an improvement in device-independent randomness generation for a photonic Bell test with a probabilistic source and for the triangle network.

 \end{abstract}

\maketitle

\section{Introduction}

Bell nonlocality is now widely recognized as a central aspect of quantum theory, playing a vital role not only in deepening our understanding of the quantum world but also in paving the way for advanced applications in quantum information processing, leading to the device-independent paradigm \cite{review}. In his seminal work \cite{bell1964einstein}, Bell demonstrated that two distant observers performing local measurements on a shared entangled state can establish strong correlations that cannot be explained by any physical theory satisfying a natural notion of locality, respected in classical physics. Such Bell nonlocal quantum correlations have been experimentally demonstrated \cite{Hensen}, and are the basis for several applications, most notably within the device-independent framework \cite{Acin2007,colbeck2009quantum,pironio2010random}. 

In recent years, the effect of quantum Bell nonlocality has been investigated in the context of quantum networks where several independent sources distribute quantum resources (e.g. entangled states) to the observers, see e.g. \cite{Branciard2010,Fritz_2012,Chaves2015,Gisin_2019, Renou_2019,boreiri2023towards,Tavakoli_Review}. The network setting allows for novel forms of Bell nonlocality, in particular in scenarios with fixed measurement settings (i.e. without requiring the observers to actively change their measurement setting) \cite{Fritz_2012,branciard2012bilocal}. In this scenario, Bell nonlocality can be used to certify the combination of entangled states and joint entangled measurements \cite{Renou_2019,renou2022network,Supic2022,Sekatski2023}, as well as for generating randomness (without the need for input randomness). So far, these questions have been mostly discussed in an idealized setting (e.g. considering deterministic sources of entanglement), and characterizing Bell nonlocality in more realistic networks remains challenging \cite{wolfe2019inflation,krivachy_neural_2020,wolfe2021inflation,boreiri2023noise}. This is indeed central towards any future experiments or applications.

In the present work, we discuss Bell nonlocality in networks with unreliable sources. This means that each source in the network, in each round of the Bell test, has a certain probability of failing, e.g. not sending out any particles. In case of such a failure, the observers connected to the source will report a failure (no-click) outcome. Dealing with such inconclusive outcomes in the context of Bell nonlocality is not trivial. In general, the naive strategy of simply discarding these inconclusive events and analyzing the nonlocality of the post-selected data is not valid. Indeed, this procedure can open the so-called detection loophole, where a local model can reproduce the (seemingly nonlocal) post-selected data by exploiting the inconclusive events \cite{pearle1970,review}. This problem is related to the fact that Bell nonlocality is predicated on certain assumptions of causal independence. In the traditional Bell scenario, the choice of measurement setting must be independent of the source of entanglement, while in network nonlocality (where measurement settings are no longer necessary), one must ensure that the different sources are independent, in accordance with the network topology. From this perspective, the detection loophole arises when the post-selection breaks the causal independence relations. In this case, the post-selected data no longer comply with the network topology, hence invalidating the proof of Bell nonlocality.

Nevertheless, here we prove that in the failing-source scenario, post-selection of inconclusive events is in fact harmless, and does not open any loophole. More precisely, we show that the correlations observed in this scenario self-test the underlying failing source model, i.e. the latter is the only quantum model compatible with the observed data. In this model, the occurrence of failure is decided at the source. Therefore post-selection does not compromise the assumption of source independence. After proving our general result, we illustrate its relevance with two examples. In particular, we apply our result to the standard photonic CHSH Bell test (with a probabilistic SPDC source of entanglement, and demonstrate its usefulness for increasing the rate of device-independent randomness generation.

From a more technical perspective, the main contributions of our work are the following. First, we define the concept of fair-sampling measurements in the context of quantum networks, following the approach of \cite{berry2010fair,orsucci2020post}. In turn, we show that whenever all measurements in the network are fair-sampling, then post-selection is harmless. Finally, to prove our main result, we prove a rigidity (self-test) statement for all correlations saturating the Finner inequality \cite{finner1992generalization,renou2019limits}. Specifically, this rigidity property guarantees that all measurements are fair-sampling.

Finally, we discuss the connection of our work with the effect of topologically robust network Bell nonlocality \cite{boreiri2024topologically} and conclude with open questions.

\section{Unreliable quantum networks and the fair sampling assumption}

Consider a network with $N$ sources $S_i$ and $M$ parties $A_j$. The network can be represented by a bipartite graph where the connections between the sources and the parties are specified by the $N\times M$ adjacency matrix $E$ taking the values $E_{ij}=1$ (connection) $E_{ij}=0$ (no connection). Each party $A_j$ performs a measurement on the systems received from the connected sources in order to produce an outcome $a_j$. In principle, before doing so it can also sample a measurement input $x_j$ from a local source of randomness. However, any such network with inputs can be mapped into a ``dressed'' network without inputs by adding a source of randomness and an extra party announcing $x_j$ (see the example in the left panel of Fig.~\ref{fig:spdc_plot}). Thus, without loss of generality, we focus here on networks without inputs. The set of possible probability distributions of outputs 
${\rm P}(a_1, a_2, \dots, a_M)\nonumber$
for a given network depends on the underlying physical theory governing the experiment. Classically (local models), a source distributes a random variable, which are read out by the measurements at the parties to produce outputs. In quantum theory, each source $S_i$ sends a quantum system $Q_{j}^{(i)}$ to the connected party $A_j$. All systems sent by a source $S_i$ are denoted by the composite system $\bm Q^{(i)} =\CP_{j| E_{ij}=1} Q_{j}^{(i)}$. The state prepared by the source $S_i$ admits a purification
$\ket{\Psi^{(i)}}_{\bm Q^{(i)}E^{(i)}}$, where $E^{(i)}$ is a purifying system potentially controlled by an eavesdropper. In turn, the measurement performed by each party are given by a POVM $\{M^{a_j}_{A_j}\}$ acting on the tensor product Hilbert space for all systems received by the party $A_j$.

\begin{figure}
    \centering
    \includegraphics[width=\columnwidth]{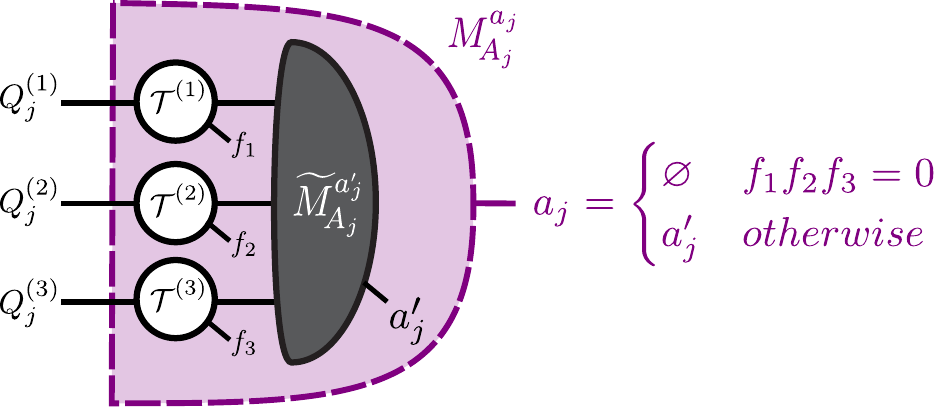}
    \centering
    \caption{A measurement $M_{A_j}^{a_j}$ is fair-sampling if it can be decomposed as shown here. First, a filter is applied on each input system, and the outcome is inconclusive $\varnothing$ if any of the filters is unsuccessful (flag output $f_i=0$). When all filters are successful, a (always) conclusive measurement $\widetilde{M}_{A_j}^{a_j}$ is performed on the filter's output $(\bigotimes_i \mathcal{T}^{(i)})[\bullet]$.}
    \label{fig:FS}
\end{figure}

To discuss networks with unreliable elements (such as failing sources), we add to each measurement an inconclusive outcome, $a_j=\varnothing$, which corresponds to a failure (no-click) event reported by party $A_j$. In turn, let $O_j^{\checkmark}$ denote the set of conclusive outputs produced by $A_j$. In general, as mentioned above, such inconclusive events cannot be simply discarded. Nevertheless, this can be done when the measurements satisfy a so-called fair-sampling property~\cite{berry2010fair,orsucci2020post}, which now formalize in the context of quantum networks. Loosely speaking, a measurement is fair-sampling if it can be performed in two steps (see Fig. \ref{fig:FS}): first, each subsystem is filtered (via a CP map), and then an ideal (always conclusive) measurement is applied.

\begin{definition}(Fair-sampling) We say that the measurement $M_{A_j}^{a_j}$ satisfies the \emph{fair-sampling} property if it admits the decomposition 
\begin{equation}\label{eq:fairsampling}\begin{split}
M_{A_j}^{a_j} &= \big(\!\bigotimes_{i|E_{ij}=1} \mathcal{T}_{Q_j^{(i)}}^{(i)}\big)^\ast \,[\widetilde M_{A_j}^{a_j}]
\quad {\rm for} \quad a_j \neq \varnothing,
\\M_{A_j}^{\varnothing} & = \id_{A_j} - \sum_{a_j\neq \varnothing} M_{A_j}^{a_j},
\end{split}
\end{equation}
where $\mathcal{T}_{Q_j^{(i)}}^{(i)}$ is a local filter (CP map) acting on the system $Q_j^{(i)}$, and $^\ast$ denotes the adjoint map, and $\widetilde M_{A_j}^{a_j}$ is a POVM which only has conclusive outcomes, as in Fig. \ref{fig:FS}.
\end{definition}

A particular case of interest is when the measurement acts on two input systems, one of which being classical and representing the input setting (as in standard Bell tests). Here we recover the previous definition of fair-sampling \cite{orsucci2020post}. Next, we derive two properties of fair-sampling measurements.

\begin{proposition}\label{propos1} The measurement $M_{A_j}^{a_j}$  is fair-sampling if and only if the (coarse-grained) POVM element corresponding to all conclusive outcomes admits the following decomposition
\be\label{eq:fairsampling2}
M_{A_j}^\checkmark  := \sum_{a_j \neq \varnothing} M_{A_j}^{a_j} = \bigotimes_{i|E_{ij}=1}  T^{(i)}_{Q_j^{(i)}} ,
\ee
with Hermitian $ 0 \preceq T^{(i)}_{Q_j^{(i)}} \preceq \id$, i.e. it is a product on all the systems it acts upon.
\end{proposition}

\begin{proposition}(Fair-sampling and harmless post-selection) \label{thrm: FS} If all measurements in the network are fair-sampling, i.e. satisfy property~\eqref{eq:fairsampling}, the post-selected distribution 
\be\label{eq: P post-selected}
     \widetilde{\rm P}(a_1, \dots, a_M)={\rm P}(a_1, \dots, a_M|a_1, \dots, a_M\neq \varnothing)
\ee
admits a quantum realization on the same network, with the states given by
\be
\ket{\widetilde{\Psi}^{(i)}}_{\bm Q^{(i)}E^{(i)}} \propto \left(\bigotimes_{j|E_{ij}=1} \! \!  \sqrt{T^{(i)}_{Q_j^{(i)}} } \right) \ket{\Psi^{(i)}}_{\bm Q^{(i)}E^{(i)}}
\ee
and the measurements given by $\{ \widetilde M_{A_j}^{a_j}\}_{a_j \in O_j^\checkmark}$ with
\be
\widetilde M_{A_j}^{a_j}= (M_{A_j}^\checkmark)^{(-1/2)} M_{A_j}^{a_j} (M_{A_j}^\checkmark)^{(-1/2)},
\ee
where $M^{(-1)}$ denotes the pseudo-inverse.
\end{proposition}

The proofs are rather straightforward and given in appendix~\ref{app: Theorem1}. The main idea is to notice that when all measurements are fair-sampling, all local filters can be moved to the sources. Hence the occurrence of failures cannot modify the network topology, and does not compromise the assumption of source independence.

In practice, it can be useful to make the fair-sampling assumption for certain tasks, for example in a non-adversarial scenario where the physics of the setup (detectors and sources) is well understood and characterized. In this case, post-selection can be safely performed without introducing any loophole, as follows from Proposition~\ref{thrm: FS}. Nevertheless, it is important to realize that when making the fair-sampling assumption, the analysis ceases to be fully device-independent. In particular, in an adversarial scenario or when working with untrusted devices, this opens the door to loopholes and potential attacks on the hardware implementation \cite{gerhardt2011experimentally,lydersen2010hacking}. Note that the fair-sampling assumption may also fail in setups that are not adversarial, and may a priori seem harmless \cite{sekatski2010cloning,pomarico2011experimental}. 


\section{Loophole free post-selection via Finner self-testing}

 Remarkably, as we show now, there exist cases where the fair-sampling property can in fact be guaranteed from the observed correlations alone. This is possible due to a self-testing property: the observed data allows one to uniquely identify the underlying quantum model.

\begin{figure}[t]
    \centering
    \includegraphics[width=\linewidth]{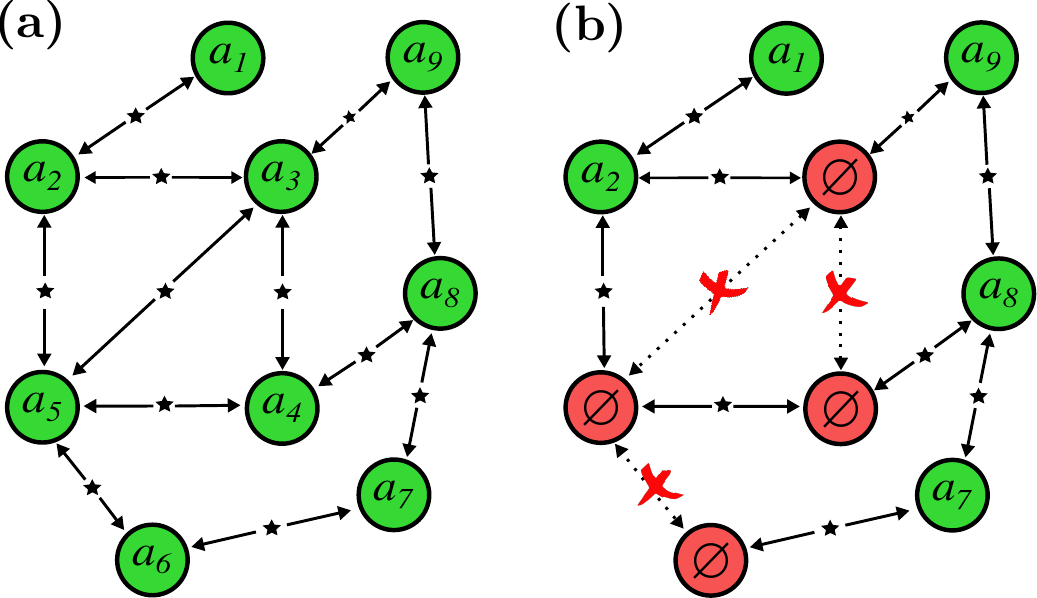}
    \caption{{\bf (a)} A quantum network with bipartite sources. Each party produces an output $a_j$ by measuring the quantum systems received from the connected sources. {\bf (b)} In the failing-source model, each source may fail with a certain probability, in which case the parties connected to it will output an inconclusive outcome $\varnothing$.} 
    \label{fig: graph failing}
\end{figure}

Here, we restrict our analysis to networks with bipartite sources, which can be represented by a graph (Fig.~\ref{fig: graph failing}), where each vertex represents a party and each edge represents a source connecting two parties. Here, the failing source model corresponds to the independent-edge (or percolation) model commonly used in network theory. Each source $S_i$ fails with a certain probability $e^{(i)}$, in which case both parties connected to it report an inconclusive outcome $\varnothing$. Conversely, a party $A_j$ outputs a conclusive outcome $a_j \in O_j^{\checkmark}$ iff neither of the sources connected to it failed. In this model, the probabilities of conclusive outcomes satisfy
$\P[a_1,\dots,a_M\neq\varnothing] =\prod_{i=1}^N (1-e^{(i)})$ and $\prod_{j=1}^M \P[a_j\neq\varnothing] =\prod_{i=1}^N (1-e^{(i)})^2$. Hence, the Finner inequality, given by \cite{finner1992generalization,renou2019limits} 
\be\label{eq:finner_eq}
\P[a_1,\dots,a_M\neq\varnothing] \leq \sqrt{\prod_{j=1}^M \P[a_j\neq \varnothing]},
\ee
is saturated here (there is equality).
Remarkably, the converse is also true: equality in Eq~\eqref{eq:finner_eq} can only be attained by a failing source model. This can be formalized as follows.

\begin{theorem}(Quantum Finner Rigidity) \label{thm: finner st} Consider a quantum network with bipartite sources $S_i$ and parties $A_j$ producing the outputs $a_{j}\in O_j^{\checkmark}\cup\{\varnothing\}$, where a source is connected to a party if $E_{ij}=1$. 
Then, the equality in Eq.~\eqref{eq:finner_eq} holds (saturation of the Finner inequality) 
if and only if, the states are of the form

\begin{align}\label{eq: state rigid}
\ket{\Psi^{(i)}}_{Q_j^{(i)}Q_{j'}^{(i)}E^{(i)}} \!\! &= \sqrt{1-e^{(i)}} \ket{11}_{{F}^{(i)}_j{F}^{(i)}_{j'}}\ket{\widetilde{\Psi}^{(i)}}_{{ \widetilde Q}^{(i)}_j { \widetilde Q}^{(i)}_{j'}E^{(i)}}\nonumber  \\
&\!\!\!+\sqrt{e^{(i)}} \ket{00}_{{F}^{(i)}_j{F}^{(i)}_{j'}}\ket{\Theta^{(i)}}_{{ \widetilde Q}^{(i)}_j { \widetilde Q}^{(i)}_{j'}E^{(i)}}
\end{align}
with $1-e^{(i)}= \frac{ \P[a_j\neq\varnothing]\, \P[a_{j'}\neq\varnothing]}{\P [a_j,a_j'\neq \varnothing]}
$ and $j,j'$ label the two parties connected to the source $S_i$,
and the measurements are of the form 
\begin{align} \label{eq: meas rigid}
M_{A_j}^{\checkmark}=\sum_{a_j \neq \varnothing} M_{A_j}^{a_j}= \bigotimes_{i|E_{ij}=1} \left(\ketbra{1}_{{ F}_{j}^{(i)}}\otimes \id_{\widetilde{Q}_{j}^{(i)}}\right).
\end{align}
Here, each system $Q_{j}^{(i)}=F_{j}^{(i)}\widetilde Q_{j}^{(i)}$ (sent by the source $S_i$ to the party $A_j$) is decomposed in the flag qubit $F_{j}^{(i)}$ and the remaining degrees of freedom $\widetilde Q_{j}^{(i)}$, $E^{(i)}$ is the system purifying the source. 
\end{theorem}






The full proof of the theorem is given in Appendix \ref{app: proof thrm 1}. The idea is to combine two previous results on the Finner inequality. First, a rigidity result has been proven for classical models~\cite{finner1992generalization}. Second, for networks with bipartite sources, Finner's inequality has been proven to hold for quantum models \cite{renou2019limits}. Following this proof, we show that if the Finner inequality is saturated, then the underlying quantum model is essentially classical: the measurement operators must be diagonal in the Schmidt basis of the states. This allows us to lift the rigidity property for classical networks to the quantum case.

Theorem~\ref{thm: finner st} states that any correlations ${\rm P}(a_1, \dots, a_M)$ saturating the Finner inequality self-test the quantum model responsible for the occurrence of the failure outcomes to be the failing-source model. In particular, by comparing Eqs.~\eqref{eq: meas rigid} and \eqref{eq:fairsampling} one sees that such correlations \emph{guarantee} that all measurements in the network are fair-sampling. 
From Proposition~\ref{thrm: FS} it follows that the post-selected correlations $\widetilde {\rm P}(a_1, \dots, a_M)$ of Eq~\eqref{eq: P post-selected} remain consistent with the network topology. In fact, this can also be directly deduced from Proposition~\ref{thm: finner st}. Indeed, accordingly to Eq.~\eqref{eq: meas rigid}, each party $A_j$ outputs a conclusive output $a_j\neq \varnothing$ iff all the flag qubits $F^{(i)}_{j}$ it received from the neighboring sources are in the state $\ket{1}$. In turn, from Eq.~\eqref{eq: state rigid} it follows that conditional on all the parties reporting a conclusive outcome, each source $S_i$  must have prepared the state $\ket{11,\widetilde{\Psi}^{(i)}_\checkmark}_{Q_{j}^i Q_{j'}^i E^{(i)}}$. Hence, if the Finner inequality is saturated post-selection does not break the source independence assumption, as summarized by the following corollary.


\begin{corollary} \label{corrolary}
     Consider a quantum network with bipartite sources, where the outputs observed by the parties include the failure events $a_j\in O_j^{\checkmark}\cup\{\varnothing\}$. If the observed correlations ${\rm P}(a_1, \dots, a_M)$ saturate the Finner inequality ~\eqref{eq:finner_eq}, then all the measurements must be fair-sampling, and the post-selected distribution $\widetilde{\rm P}(a_1, \dots, a_M)$ in Eq.~\eqref{eq: P post-selected} admits a quantum model on the same network. 
\end{corollary}

We now illustrate the usefulness of these results with two examples, where Bell nonlocality is used to certify randomness.

\subsection{CHSH test with SPDC source}

\begin{figure*}[t]
    \centering
    \includegraphics[width=1.1\columnwidth]{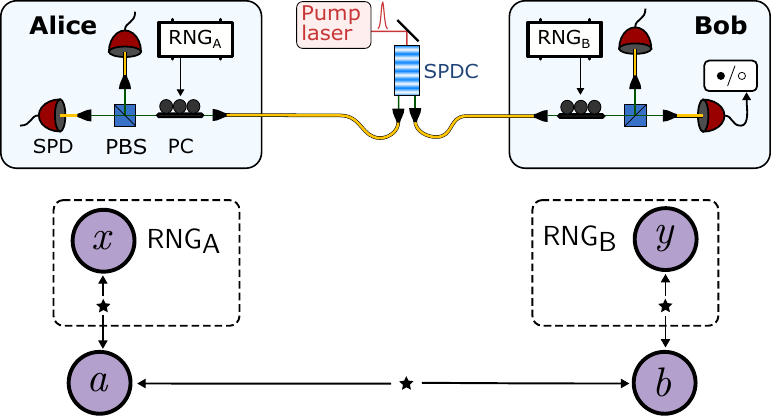}
    \includegraphics[height=0.57\columnwidth]{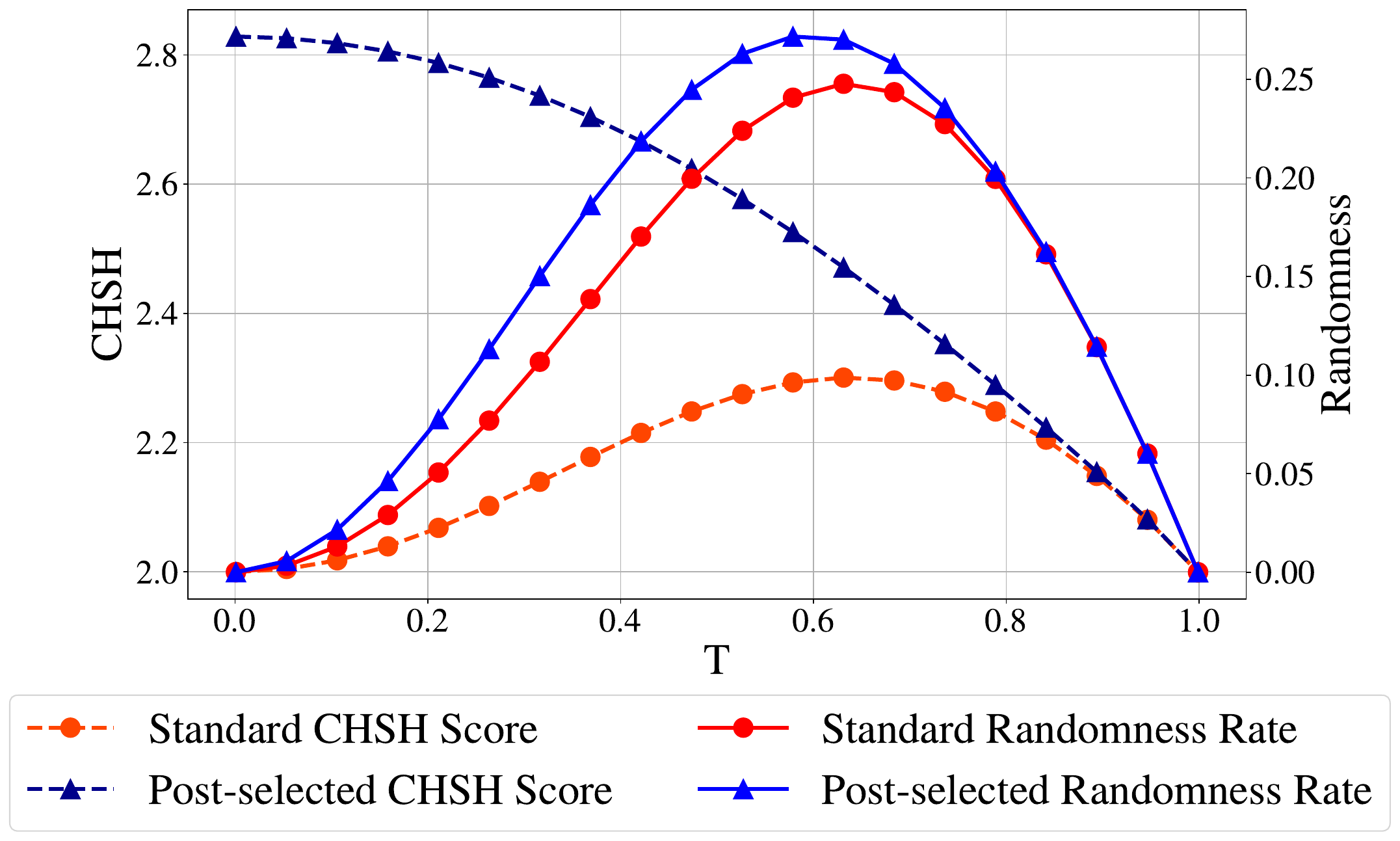}
    \caption{{\bf(Top left)} Sketch of an experimental setup for testing the CHSH Bell inequality with an SPDC source of polarization-entangled photons and single-photon detectors. {\bf(Bottom left)} The three-source network describing the experiment. {\bf(Right)} CHSH score (left scale, dashed) and randomness rate (right scale, full) obtained from the bare (red) and post-selected (blue) data, via the Corollary~\ref{corrolary}, as the function of the pump parameters $T=T_1=T_2$ of the source.}
    \label{fig:spdc_plot}
\end{figure*}

To start, we consider a very concrete example of the bipartite Bell test (see Fig.~\ref{fig:spdc_plot}, where an optical source based on spontaneous parametric down-conversion (SPDC) distributes polarisation entangled photons pairs to Alice and Bob. Their measurement settings are generated via local random number generators (RNGs). Here we represent this setup as a network with three independent sources, as in Fig. 3. 


The state prepared by the SPDC source is of the form  
\begin{equation}
\ket{\Psi} = \sqrt{\mathcal{N}} \exp(T_1 a_h^\dag b_h^\dag + T_2 a_v^\dag b_v^\dag) \ket{\bm 0} \label{eq: SPDC state gen}
\end{equation}
where $\mathcal{N} = (1+T_1^2)(1+T_2^2)$, $a_{h(v)}$ and $b_{h(v)}$ are the photonic modes sent to Alice and Bob respectively, $\ket{\bm 0}$ is the vacuum state of all the involved modes, and $T_1$ and $T_2$ parametrize the intensity of the pump laser. The SPDC source is probabilistic by nature. It creates no photons with probability $p_0=\mathcal{N}$, a single photon pair with probability $p_1=\mathcal{N}(T_1^2+T_2^2)$, and multiple pairs with the remaining probability.

Alice and Bob measure locally the polarization of the photons. After a polarization rotation controlled by the measurement setting, orthogonally polarized photons are separated via a polarizing beamsplitter (PBS) and measured via non-photon resolving detectors (SPD).  
Such detectors are binary: they either click  $D_\bullet = \id -\ketbra{0}$ or not $D_\circ= \ketbra{0}$. Hence, each measurements ($E_{a|x}$ and $E_{b|y}$) is described by a four-outcome POVM with outputs $a,b \in\{\circ\circ,\bullet\circ,\circ\bullet,\bullet\bullet\}$. The local outcome is conclusive when only one of the two SPDs clicks, e.g. $a= \circ\bullet$  or $a=\bullet\circ$, and inconclusive otherwise. Note that when the source generates a single photon pair, outcomes are conclusive, whereas if the source emits no or multiple pairs, inconclusive outcome will occur.

This setup is commonly used to test the CHSH Bell inequality \cite{CHSH} and for device-independent randomness generation \cite{Christensen2013,Bierhorst2018}. As this test requires binary outputs, it is common to bin inconclusive outcomes together with a conclusive one, e.g. a=0 for $\circ\circ$, $\bullet\bullet$, and $\circ\bullet$, and a=1 for $\bullet\circ$ \cite{vivoli2015comparing}. Lower bounds on the outcome randomness, given by the conditional von Neumann entropy, are obtained from the CHSH score~\cite{pironio2009device}. Traditionally, this operation is performed using the raw data. 

Using the above results, we can analyze the data differently. Indeed, both parties observe the local outcome $\circ\circ$ only when the source produces vacuum (no pair), and these events are perfectly correlated. Labeling these as failure outcomes $\circ\circ \mapsto \varnothing$ we recover the failing-source model described above. Then, corollary ~\ref{corrolary} guarantees that these inconclusive events can be safely discarded. We then evaluate the CHSH value and the randomness bound on the post-selected data. Taking into account the post-selection rate, we plot in Fig.~\ref{fig:spdc_plot} the randomness rate as a function of the pump intensity. Interestingly, we obtain a small but non-negligible improvement over the standard method where no data is discarded. All details can be found in appendix~\ref{app: SPDC}. 

Finally, note that here we have neglected other imperfections in the setup (such as channel loss), and restricted to the case of collective attacks. Applying a similar analysis to practical setups would require generalizing the Theorem \ref{thm: finner st} to correlations that do not saturate the Finner inequality exactly.

\subsection{Nonlocality and Randomness in the triangle network}

Next, we consider the triangle network scenario \cite{Fritz_2012,Renou_2019}, consisting of three parties -- A, B, and C — connected pairwise by bipartite sources labeled with $\alpha, \beta,\gamma$. Each party receives two systems from the neighboring sources and produces an output 
$a, b, c$. Overall, there are six systems involved, labeled as $Q^{(i)}_j$ where $i \in \{\alpha, \beta,\gamma\}$ denotes the source generating it and $j\in \{ A,B,C\}$ identifies the connected party receiving the system. 
As a concrete example, we consider the quantum distribution discussed in \cite{Renou_2019} for which a bound on the local output randomness can be derived~\cite{Sekatski2023}. 


In the failing source model, the analysis of Refs \cite{Renou_2019,Sekatski2023} no longer works. Nevertheless, we will see that Bell nonlocality and randomness can still be certified using our approach. Let us assume that each source fails with probability $e^{(i)}$, leading the neighboring parties to report a failure outcome. First, we observe that the resulting correlations will saturate the Finner inequality~\eqref{eq:finner_eq}, and by Theorem~\ref{thm: finner st} the states prepared by the sources must be of the form in Eq.~\eqref{eq: state rigid}. Hence, all the outputs are conclusive iff the sources prepare the flag qubits in the states $\ket{11}_{ F^{(i)}_j F^{(i)}_{j'}}$, and the post-selected statistics $\widetilde{\rm P}(a,b,c)$ are consistent with the triangle network. Hence we can apply previous results where the states $\ket{\widetilde{\Psi}^{(i)}}_{{ \widetilde Q}^{(i)}_j { \widetilde Q}^{(i)}_{j'}E^{(i)}}$ can be partially self-tested \cite{Sekatski2023}. Following this proof, we can thus certify Bell nonlocality and derive a bound on the output randomness. For example, if in the ideal scenario (perfect source) we have $H({\rm A}|E) \geq L$, with $\rm A$ denoting the classical register containing the output $a$ and $H$ being the von Neuman entropy, then in the failing source model, we obtain the modified bound
\begin{equation}
H({\rm A}|E) \geq (1 - e^{(\beta)})(1 - e^{(\gamma)})~L.
\end{equation}
It is worth noting that this bound is stronger than the one derived from directly using post-selected data, which would yield a coefficient of $(1 - e^{(\alpha)})(1 - e^{(\beta)})(1 - e^{(\gamma)})$, accounting for non-failure of all of the sources. This is coming from applying the self-testing directly as we need only the neighbouring sources $\alpha$ and $\gamma$ to not fail, for the output of $A$ to be conclusive. All details can be found in appendix~\ref{app: RGB4}. \\





\section{Discussion}


We discussed Bell nonlocality in quantum networks with unreliable sources. Our main result is a condition on the observed data which ensures that inconclusive events can be safely discarded, without introducing any loophole. Our work contributes to understand the robustness of Bell nonlocality in quantum networks.

To do so, we have introduced the notion of fair-sampling measurements for networks, which may be of independent interest and facilitate experimental implementations. 

Our work is also interesting from the perspective of testing Bell nonlocality in networks with varying topology. Here, in each round of the test, the topology varies depending on which sources failed (see Fig. 2). This is complementary to the results of \cite{boreiri2024topologically} where the topology is partially characterized, but fixed throughout the test. 

Our results may also find applications, e.g. in randomness generation as we illustrated with examples. For practical purposes, one would need to include the effect of other noise sources, in particular channel loss. Technically, this would require an approximate version of our Finner self-testing result, along the lines of \cite{ellis2015geometric,boreiri2023noise}. Finally, one could explore the extension of the Finner inequality and self-testing for more complex networks, involving multipartite sources.

\emph{Acknowledgements.---} We thank Joey Marcellino and Patrik Caspar for their assistance. We acknowledge financial support from the Swiss National Science Foundation (project 192244) and the by the Swiss Secretariat for Education, Research and Innovation (SERI) under contract number UeM019-3.

\onecolumngrid
\appendix

\setcounter{theorem}{0}

\newpage
\section{Proof of Propositions 1 and 2}

\begin{proposition} The following properties of the measurement $M_{A_j}^{a_j}$  are equivalent
\begin{itemize}
    \item[(i)] Fair-sampling definition: the POVM elements admit a decomposition
    \begin{equation}\begin{split}
M_{A_j}^{a_j} &= \big(\bigotimes_{i|E_{ij}=1} \mathcal{T}_{Q_j^{(i)}}^{(i)\ast}\big)[\widetilde M_{A_j}^{a_j}]
\quad {\rm for} \quad a_j \neq \varnothing,
\\M_{A_j}^{\varnothing} & = \id_{A_j} - \sum_{a_j\neq \varnothing} M_{A_j}^{a_j},
\end{split}
\end{equation}
where $\mathcal{T}_{Q_j^{(i)}}^{(i)}$ are CP maps acting on the systems $Q_j^{(i)}$, and $\ast$ denotes the adjoint, and $\widetilde M_{A_j}^{a_j}$ is a POVM which only has conclusive outcomes.
\item[(ii)] The POVM element corresponding to all conclusive outcomes satisfies
\be
M_{A_j}^\checkmark  := \sum_{a_j \neq \varnothing} M_{A_j}^{a_j} = \bigotimes_{i|E_{ij}=1}  T^{(i)}_{Q_j^{(i)}}
\ee
for some Hermitian $ 0 \preceq T^{(i)}_{Q_j^{(i)}} \preceq \id$.
\end{itemize}
\end{proposition}
\begin{proof}
    The implication $(i)\implies(ii)$ is straightforward. The measurement $\widetilde M_{A_j}^{a_j}$ only has conclusive outcomes, therefore  $\sum_{a_j \neq \varnothing}\widetilde M_{A_j}^{a_j} = \id_{A_j} = \bigotimes_{i|E_{ij}=1}  \id_{Q_j^{(i)}}$ and 
    \begin{align}
        M_{A_j}^\checkmark  &= \sum_{a_j \neq \varnothing} M_{A_j}^{a_j} = \sum_{a_j \neq \varnothing} \big(\bigotimes_{i|E_{ij}=1} \mathcal{T}_{Q_j^{(i)}}^{(i)\ast}\big)[\widetilde M_{A_j}^{a_j}] \\
       &= \big(\bigotimes_{i|E_{ij}=1} \mathcal{T}_{Q_j^{(i)}}^{(i)\ast}\big)[\id_{A_j}]
       \\ &=\bigotimes_{i|E_{ij}=1} \mathcal{T}_{Q_j^{(i)}}^{(i)\ast}[\id_{Q_j^{(i)}}] \\
       & := \bigotimes_{i|E_{ij}=1}  T^{(i)}_{Q_j^{(i)}}.
    \end{align}

To show the implication $(ii)\implies(i)$, we will now drop the superscript of $T_{Q_j^{(i)}}^{(i)}$ to avoid cumbersome expression. For each positive semi-definite operator $T_{Q_j^{(i)}}$  define its pseudo-inverse $T_{Q_j^{(i)}}^{(-1)} \succeq 0$ as well as the projector on its support $\Pi^{T}_{Q_j^{(i)}}$, such that 
$T_{Q_j^{(i)}}^{(-1)}T_{Q_j^{(i)}} =T_{Q_j^{(i)}} T_{Q_j^{(i)}}^{(-1)} = \Pi^{T}_{Q_j^{(i)}}$. Analogously we define their tensor products
\be
(M_{A_j}^{\checkmark})^{(-1)} = \bigotimes_{i|E_{ij}=1}  T_{Q_j^{(i)}}^{(-1)} \qquad \text{and} \qquad  \Pi_{A_j}^{T} = \bigotimes_{i|E_{ij}=1}  \Pi^{T}_{Q_j^{(i)}}
\ee
Now consider the following operators 
\be
\widetilde M_{A_j}^{a_j} = (M_{A_j}^{\checkmark})^{(-1/2)} M_{A_j}^{a_j} (M_{A_j}^{\checkmark})^{(-1/2)}, 
\ee
they are positive semi-definite, and satisfy
\be
\sum_{a_j\in O_j^\checkmark} \widetilde M_{A_j}^{a_j} = 
(M_{A_j}^{\checkmark})^{(-1/2)} \sum_{a_j\in O_j^\checkmark}M_{A_j}^{a_j}(M_{A_j}^{\checkmark})^{(-1/2)} 
=(M_{A_j}^{\checkmark})^{(-1/2)} M_{A_j}^{\checkmark} (M_{A_j}^{\checkmark})^{(-1/2)} = \Pi_{A_j}^T.
\ee
Hence $\{\widetilde M_{A_j}^{a_j} \}_{a_j \in O_j^\checkmark}$ for a POVM on the subspace defined by $\Pi_{A_j}^T$. If $\Pi_{A_j}^T \neq \id_{A_j}$ one can always extend the POVM $\{\widetilde M_{A_j}^{a_j} \}_{a_j \in O_j^\checkmark}$ to act on the whole Hilbert space of the system $A_j$ by associating $ \id_{A_j}-\Pi_{A_j}^T $ to any conclusive outcome (the state is never found there anyway).

By defining the (L\"uders) quantum instrument $\mathcal{I}_{A_j}= \{\mathcal{T}_{A_j} , \mathcal{F}_{A_j}\}$ composed of two CP maps given by
\be
\mathcal{T}_{A_j} (\bullet) = \sqrt{M_{A_j}^\checkmark} \bullet \sqrt{M_{A_j}^\checkmark } \qquad \mathcal{F}_{A_j} (\bullet) = \sqrt{M_{A_j}^\varnothing} \bullet \sqrt{M_{A_j}^\varnothing},
\ee
it is always possible to decompose our POVM as
\begin{align}
M_{A_j}^{a_j} &= \mathcal{T}_{A_j}^*(\widetilde M_{A_j}^{a_j})\quad \text{for} \quad a_j \in O_j^{\checkmark}\\ 
M_{A_j}^{\varnothing} &= \mathcal{F}_{A_j}^*(\id_{A_j}).
\end{align}
This is a two-step process. First, the instrument $\mathcal{I}_{A_j}$  verifies if a failure outcome must be reported ($\mathcal{F}_{A_j}$) or otherwise ($\mathcal{T}_{A_j}$) a conclusive outcome must be reported. In the later case the instrument outputs the quantum system in the subspace $\Pi^{T}_{Q_j^{(i)}}$ and the POVM $\{\widetilde M_{A_j}^{a_j} \}_{a_j \in O_j^\checkmark}$ is performed on it to determine which conclusive outcome will be output. Furthermore, by assumption we know that the CP map $\mathcal{T}_{A_j}= \bigotimes_{i|E_{ij}=1} \mathcal{T}_{Q_{j}^{(i)}}^{(i)}$ is product, with $\mathcal{T}_{Q_{j}^{(i)}}^{(i)}(\bullet) := \sqrt{T_{Q_{j}^{(i)}}^{(i)}}\bullet \sqrt{T_{Q_{j}^{(i)}}^{(i)}}$. Hence,
\begin{align}
M_{A_j}^{a_j} &= \mathcal{T}_{A_j}^*(\widetilde M_{A_j}^{a_j})= \big(\bigotimes_{i|E_{ij}=1} \mathcal{T}_{Q_{j}^{(i)}}^{(i)\ast} \big)[\widetilde M_{A_j}^{a_j}] \quad \text{for} \quad a_j \in O_j^{\checkmark}\\ 
M_{A_j}^{\varnothing} &= \mathcal{F}_{A_j}^*(\id_{A_j})= \id_{A_j}-\sum_{a_j\neq\varnothing} M_{A_j}^{a_j} 
\end{align}
\end{proof}

\begin{proposition} If all measurements  in the network are fair-sampling the post-selected distribution 
$     \widetilde{\rm P}(a_1, \dots, a_M)={\rm P}(a_1, \dots, a_M|a_1, \dots, a_M\neq \varnothing)$
admits a quantum realization on the same network, with the states given by
\be
\ket{\widetilde{\Psi}^{(i)}}_{\bm Q^{(i)}E^{(i)}} \propto \left(\bigotimes_{j|E_{ij}=1} \! \!  \sqrt{T^{(i)}_{Q_j^{(i)}} } \right) \ket{\Psi^{(i)}}_{\bm Q^{(i)}E^{(i)}}
\ee
and the measurements given by $\{ \widetilde M_{A_j}^{a_j}\}_{a_j \in O_j^\checkmark}$ with
\be
\widetilde M_{A_j}^{a_j}= (M_{A_j}^\checkmark)^{(-1/2)} M_{A_j}^{a_j} (M_{A_j}^\checkmark)^{(-1/2)},
\ee
where $M^{(-1)}$ denotes the pseudo-inverse.
\end{proposition}

\label{app: Theorem1}
Consider a quantum network with $N$ sources $S_i$ and $M$ parties $A_j$ producing the outputs $a_j\in O_j^\checkmark \cup\{\varnothing\}$. A source is connected to a party if $E_{ij}=1$, in which case it sends a quantum system $Q_j^{(i)}$. By assumption, each measurement in the network satisfies the fair-sampling property, i.e. for conclusive outcomes $a_j\neq \varnothing$
\be
M_{A_j}^{a_j} = \big(\bigotimes_{i|E_{ij}=1} \mathcal{T}_{Q_{j}^{(i)}}^{(i)\ast} \big)[\widetilde M_{A_j}^{a_j}] = \left(\bigotimes_{j|E_{ij}=1} \! \!  \sqrt{T^{(i)}_{Q_j^{(i)}} } \right) \widetilde M_{A_j}^{a_j}\left(\bigotimes_{j|E_{ij}=1} \! \!  \sqrt{T^{(i)}_{Q_j^{(i)}} } \right)
\ee
for some Hermitian operators $0\preceq T_{Q_j^{(i)}}^{(i)}\preceq\id_{Q_j^{(i)}}$, and a measurement $\widetilde M_{A_j}^{a_j}$ that has conclusive outcomes only (i.e. $\sum_{a_j\neq \varnothing} \widetilde M_{A_j}^{a_j} = \id_{A_j}$).

Therefore, when all the outcomes are successful their occurance probability admits the following decomposition
\be
{\rm P} (a_1,\dots a_M; a_1,\dots ,a_M\neq \varnothing) = \bra{ \Psi}_{\bm Q\bm E}\left(\bigotimes_{ij|E_{ij}=1} \sqrt{T^{(i)}_{Q_j^{(i)}}} \right) (\bigotimes_{j=1}^M \widetilde M_{A_j}^{a_j}) \left(\bigotimes_{ij|E_{ij}=1} \sqrt{T^{(i)}_{Q_j^{(i)}}} \right)\ket{\Psi}_{\bm Q\bm E}
\ee
where $\ket{\Psi}_{\bm Q\bm E} = \bigotimes_{i=1}^N \ket{\Psi^{(i)}}_{\bm Q^{(i)}E^{(i)}}$ denotes the global state prepared by all the sources. Regrouping the local filters implies
\begin{align}
{\rm P} (a_1,\dots a_M; a_1,\dots ,a_M\neq \varnothing) &= \P( a_1,\dots ,a_M\neq \varnothing) \bra{\widetilde \Psi}_{\bm Q\bm E}(\bigotimes_{j=1}^M \widetilde M_{A_j}^{a_j}) \ket{\widetilde \Psi}_{\bm Q\bm E}  \qquad \text{with} \\
\ket{\widetilde \Psi}_{\bm Q\bm E} &=\bigotimes_{i=1}^N \ket{\widetilde \Psi^{(i)}}_{\bm Q^{(i)}E^{(i)}} \\
\ket{\widetilde \Psi^{(i)}}_{\bm Q^{(i)}E^{(i)}} & = \frac{\left(\bigotimes_{j|E_{ij}=1} \sqrt{T^{(i)}_{Q_j^{(i)}}} \right) \ket{\Psi^{(i)}}_{\bm Q^{(i)}E^{(i)}}}{\|\left(\bigotimes_{j|E_{ij}=1} \sqrt{T^{(i)}_{Q_j^{(i)}}} \right) \ket{\Psi^{(i)}}_{\bm Q^{(i)}E^{(i)}}\|} \\
\P( a_1,\dots ,a_M\neq \varnothing) &= \prod_{i=1}^N \|\left(\bigotimes_{j|E_{ij}=1} \sqrt{T^{(i)}_{Q_j^{(i)}}} \right) \ket{\Psi^{(i)}}_{\bm Q^{(i)}E^{(i)}}\|^2.
\end{align}
Dividing both sides by $\P( a_1,\dots ,a_M\neq \varnothing)$, we get
\begin{align}
{\rm P} (a_1,\dots a_M| a_1,\dots ,a_M\neq \varnothing) =  \bra{\widetilde \Psi}_{\bm Q\bm E}(\bigotimes_{j=1}^M \widetilde M_{A_j}^{a_j}) \ket{\widetilde \Psi}_{\bm Q\bm E}  \qquad 
\end{align}
concluding the proof.

\section{Proof of Theorem 3}
\label{app: proof thrm 1}

This appendix is dedicated to prove the following proposition from the main text. Note that the phrasing of the proposition is slightly different below, however we show that it is equivalent in section \ref{app: corrolary}.


\begin{theorem}\label{thrm: Q rigidity finner}Consider a quantum network with $N$ bipartite sources $S_i$ connected to $M$ parties $A_j$ accordingly to the $N\times M$ adjacency matrix $E_{ij}\in\{0,1\}$. Each source $S_i$ is connected to the party $A_j$ and sends it the quantum system $Q_{j}^{(i)}$ iff $E_{ij}=1$. Let $\bm Q^{(i)}=Q_{j}^{(i)}Q_{j'}^{(i)}$ where $j,j'$ label the two parties connected to the source $S_i$, and $\bm Q_j= \CP_{i|E_{ij}} Q_{j}^{(i)}$ collect all the systems received by the party $A_j$. Upon receiving these systems each source produces a binary outcome $a_j\in\{0,1\}$ by performing some POVM $M_{\bm Q_j}^{a_j}$ on the received systems. The correlations of the outputs saturate the quantum Finner inequality [cite]
\be\label{app: finner eq}
\P\big[ a_1=\dots=a_M=1\big] = \sqrt{\prod_j \P\big[a_j=1\big]} \qquad 
\ee
if and only if in the underlying quantum model 
the states prepared by each source $S_i$ admit a purification $\ket{ \Psi^{(i)}}_{\bm Q^{(i)}}$ such that
\begin{align}
\ket{ \Psi^{(i)}}_{\bm Q^{(i)}} &= \sum_{\ell_i=1}^{d_i} \lambda_{\ell_i}^{(i)}\ket{\ell_i,\ell_i}_{Q_{j}^{(i)}Q_{j'}^{(i)}} \\
M_{\bm Q_j}^1 &= \bigotimes_{i|E_{ij}=1} \Pi^1_{Q_j^{(i)}} \qquad \text{with}\qquad
\Pi^1_{Q_j^{(i)}} =  \Pi^1_{Q_{j'}^{(i)}}= \sum_{\ell_i=1}^{d_i} \chi^{(i)}(\ell_i)\ketbra{\ell_i}
\end{align}
where $\chi^{(i)}(\ell_i)\in\{0,1\}$.
In addition, for each source $S_i$, connected to the parties $A_j$ and $A_{j'}$ we define: 
\be \label{eq: pi}
e^{(i)}:=1-\sum_{\ell_i}^{d_i} (\lambda_{\ell_i}^{(i)})^2 \chi^{(i)}(\ell_i) = 1-\frac{ \P[a_j=1]\P[a_{j'}=1]}{\P [a_j =a_j'=1]}.
\ee
\end{theorem}

Note that in the failing source model, $e^{(i)}$ can be interpreted as the probability of the sources $S_i$ to fail, see the discussion in Sec.~\ref{app: corrolary}.

To prove it we proceed in a few steps. First, in sec.~\ref{app: c finner} we recall the analogous result for classical models. Second, in sec.~\ref{app: notation} we briefly introduce the necessary notation and give an a priori description of any quantum model on the network. Then, in sec.~\ref{app: q finner} following \cite{renou2019limits} we give a brief proof of the quantum Finner inequality (for binary sources). Finally, in sec.~\ref{app: final proof} we prove the theorem, to do so given the equality 
\eqref{app: finner eq} we identify which steps in the inequality proof must be tight. Then we derive a proof for the theorem by combining the classical rigidity with elementary operator algebra. In addition, in sec.~\ref{app: corrolary} we establish the connection between the different formulations of the theorem given here and in the main text.

\subsection{Classical rigidity from the Finner equality condition}
\label{app: c finner}

Consider a network $\cN$ with $M$ parties $A_1, \dots, A_M$ and $N$ sources $ S_1, \dots S_N$ specified by the $M\times N$ adjacency matrix, where a party $A_i$ is connected to the source $S_j$ if $E_{ij}=1$. We assume that each party (source) is connected to at least one source (party). Furthermore, we assume that no source is redundant, i.e. there is no pair $i \neq i'$ such that
\begin{equation}
    \bigcup_{j|E_{ij}=1} \! A_j \subset \bigcup_{j|E_{i' j}=1} \! A_j,
\end{equation}
as a redundant source $S_i$ can always be absorbed inside $S_{i' }$.


\begin{theorem} (Classic Finner inequality \cite{finner1992generalization}) Consider a network $\mathcal{N}$ with $M$ observers $A_j$. Let $\{x_1, ..., x_M\}$ be a fractional independent set of $\mathcal{N}$, i.e. weights $x_j$ attributed to each $A_j$ such that, for each source, the sum of the weights of parties connected to it is smaller or equal to 1. 
Let $f_{j}$ be any real local function of the sources $A_j$ has access to, i.e. $f_j(\{s_i: E_{ij}=1\})$. Then, any possible correlation satisfies the following inequality
\begin{equation}\label{eq:Finner}
\mean{\prod_j |f_j|} \leq \prod_j \norm{f_j}_{1/x_j},
\end{equation}
where $\norm{f}_{\alpha}= \Big(\mean{|f|^\alpha}\Big)^{1/\alpha} $ is the Schatten norm.\\

\textbf{Equality condition:} Given that the fractional independent set $\{x_j\}$ is perfect, i.e. $\sum_j {x_j = 1}$, equality in \ref{eq:Finner} holds if and only if for all $j \in [M]$ and $i: E_{ij}=1$, there exist functions $f_{j,i}$ and constants $A_{j,i}>0$ s.t:

\begin{itemize}
\item 
$|f_j(\{s_i| E_{ij}=1\})|=\prod_{i:E_{ij}=1} |f_{j,i}(s_i)|  \quad \forall j$

\item 
$A_{r,i} |f_{r,i}|^{1/x_{r}} = A_{s,i} |f_{s,i}|^{1/x_{s}}  \quad \forall i,~\text{and} ~ r,s \text{~with~} E_{ir}=E_{is}=1$
\end{itemize}

\end{theorem}

\begin{remark} In particular,
in case of $f_j(\{s_i| E_{ij}=1\})\in\{0,1\}$ being the indicator function of the output being $a_j$, we obtain:
\begin{equation}\label{app:NQFinnerProba}
P(a_1,\dots,a_M)\leq \prod_{j=1}^M \left(P_{A_{j}}(a_j)\right)^{x_j}. 
\end{equation}
In this case, given that the fractional independent set $\{x_j\}$ is perfect, i.e. $\sum_j {x_j = 1}$, equality in \ref{app:NQFinnerProba} holds if and only if there exist \emph{indicator} functions $\phi_i(s_i)\in\{0,1\}$ such that 
\begin{equation}\label{app:eq:finner-eq-thm}
f_j(\{s_i| E_{ij}=1\})=\prod_{i:E_{ij}=1} \phi_i (s_i), \quad \forall j.
\end{equation}
\end{remark}

\begin{remark}\label{remark:finner_eq_prob}
    On graphs, we can consider the fractional independent set to be $x_j=1/2 \, \forall j$, therefore the inequality would be
    \begin{equation}
    \mean{\prod_j |f_j|} \leq \prod_j \sqrt{\mean{f_j^2}}
\end{equation}
and equality holds if and only if there exist functions $h_{i} \geq 0 $ for any source $i$ and constant $C_j$ for any party $j$  s.t:

$$|f_j\{(s_i|E_{ij}=1)\}|= C_j \prod_{i:E_{ij}=1} h_{i}(s_i)   \quad \forall j$$

\end{remark}


\subsection{Notation and a priori description of quantum models}

\label{app: notation}

Recall that our network is represented by the bipartite graph where $N$ independent source $S_1,\dots,S_N$ are connected to $M$ parties $A_1,\dots, A_M$, where the $N\times M$ adjacency matrix $E_{ij}\in\{0,1\}$ indicating if the source $S_i$ is connected to the party $A_j$.

For  a quantum model each edge $E_{ij}=1$ of the graph must correspond to a quantum system $Q_j^{(i)}$ sent by the source $S_i$ to the party $A_j$. In the statement of the theorem, we have also used the following shortened notation
\be
\bm Q^{(i)} = \CP_{j|E_{ij}=1} Q_j^{(i)} \qquad \text{and} \qquad {\bm Q}_j = \CP_{i|E_{ij}=1} Q_j^{(i)},
\ee
where $\bm Q^{(i)}$ collects all the systems prepared by the source $S_i$, and ${\bm Q}_j$ collects all the quantum systems received by the party $A_j$. Any quantum model on the network is given by some density operators $\rho^{(i)}_{\bm Q^{(i)}}$ describing the states produced by the sources $S_i$, and POVMs $\{M_{\bm Q_j}^{(a_j)}\}$ describing the measurements performed by the parties $A_j$.\\

In general, the states $\rho^{(i)}_{\bm Q^{(i)}}$ prepared by the source do not have to be pure. Nevertheless by introducing an auxiliary quantum system $E^{(i)}$ for each source, thestate can be purified to
\be
\ket{\Psi^{(i)}}_{\bm Q^{(i)}E^{(i)}} \quad \text{wiht} \qquad \rho^{(i)}_{\bm Q^{(i)}}=\tr_{E^{(i)}} \ketbra{\Psi^{(i)}}_{\bm Q^{(i)}E^{(i)}}.
\ee
Hence for any quantum model on the network, there exists a purified model. To keep the notation short the purifying system $E^{(i)}$ can then be absorbed into one of the quantum systems prepared by the source. From now on we will thus reason in terms of the purified model without explicitly mentioning the system $E^{(i)}$, for simplicity of the notation. \\

Since each source in our network is bipartite, each $\bm Q^{(i)}$ contains two systems and each state $\ket{\Psi^{(i)}}_{\bm Q^{(i)}}$ admits a Schmidt decomposition
\be
\ket{\Psi^{(i)}}_{\bm Q^{(i)}} = \sum_{\ell_i=1}^{d_i} \lambda_{\ell_i}^{(i)}\ket{\ell_i,\ell_i}_{\bm Q^{(i)}},
\ee
fixing a basis for each quantum system in the network. We can now compute the probability that all parties observe 1. It is given by 
\begin{align}
\P[a_1=\dots= a_M =1] &= \bigotimes_{i=1}^N \bra{\Psi^{i}}_{\bm Q^{(i)}} \left(\bigotimes_{j=1}^M M_{\bm Q_j}^1\right)   \bigotimes_{i'=1}^N \ket{\Psi^{i'}}_{\bm Q^{(i')}} \\
&= \sum_{\bm \ell, \bm \ell'} \lambda_{\ell_1}^{(1)}\lambda_{\ell_1'}^{(1)}\dots \lambda_{\ell_N}^{(N)}\lambda_{\ell_N'}^{(N)} \bra{\bm \ell}  \left(\bigotimes_{j=1}^M M_{\bm Q_j}^1\right)  \ket{\bm \ell'}
\end{align}

with strings $\bm \ell = (\ell_1,\dots, \ell_N)$ and $\bm \ell' = (\ell_1',\dots, \ell_N')$. It will also be useful to compute the marginal state of the systems $\bm Q_j$ received by the party $A_j$, it is given by 
\be
\rho_{\bm Q_j} = \bigotimes_{i|E_{ij}=1} \left(\sum_{\ell_i=1}^{d_i} (\lambda_{\ell_i}^{(i)})^2\ketbra{\ell_i}_{Q_{j}^{(i)}} \right).
\ee
Which allows one to compute the marginal probability
\be
\P[a_j=1] = \tr \rho_{\bm Q_j} M_{\bm Q_j}^1.
\ee

\subsection{Proof of the quantum Finner inequality (binary sources) \cite{renou2019limits}}

\label{app: q finner}

In \cite{renou2019limits}, it has been shown that the Finner inequality holds in quantum theory as well, for graphs(networks with bipartite sources). Here we investigate the equality condition of the Finner inequality and prove a self-testing result in that regard.

For any quantum model, we now define a local  variable (LV) model capable of mimicking some quantum expected value $\P[a_1=\dots= a_M =1]$. First, let each source
$S_i$  uniformly sample a pair of independent integers $(\ell_i,\ell_i')$ which both have the cardinality $d_i$ equal to the Schmidt rank of the state $\ket{\Psi^i}$, hence they are distributed with ${\rm p}(\ell_i,\ell_i')=\frac{1}{d_i^2}$. After sampling these integers, the sends a copy of these values $(\ell_i,\ell_i')$ to both parties connected to it. \\

Each party $A_j$ thus receives the following strings of integers $(\bm \ell_j,\bm \ell_j')$ with
\be
\bm \ell_j =\CP_{i|E_{ij}=1} \!\ell_i \qquad \text{and} \qquad {\bm \ell}_j' =\CP_{i|E_{ij}=1} \!\ell_i'.
\ee
Both strings $\bm \ell_j$ and $\bm \ell_j'$ contain an integer label for each sources connected to the party $A_j$, and can be used to define quantum states $\ket{\bm \ell_j}_{\bm Q_j} = \bigotimes_{i|E_{ij}=1} \ket{\ell_i}_{Q_j^{(i)}}$ and $\ket{\bm \ell_j'}_{\bm Q_j} = \bigotimes_{i|E_{ij}=1} \ket{\ell_i'}_{Q_j^{(i)}}$ for the corresponding quantum systems.  The party will then output a nonegative real value 
\be\label{eq: g def}
g_j(\bm \ell_j, {\bm \ell}_j') =  \left(\prod_{i|E_{ij}=1} d_i \sqrt{  \lambda^{(i)}_{\ell_i} \lambda^{(i)}_{\ell_i'} } \right) \left|  \bra{\bm \ell_j}_{\bm Q_j} M_{\bm Q_j}^1 \ket{\bm \ell_j'}_{\bm Q_j}\right|,
\ee
computed local from the random variable $\bm \ell_j$ and  ${\bm \ell}_j'$. 

Let us now show that for the LV model we just defined the expected value of the product of all $g_j$ is equal to the probability $\P[a_1=\dots =a_M =1] $ of the quantum model. We have
\begin{align}
    \E\left [ \left| \prod_j g_j\right| \right] &= \sum_{\bm \ell,\bm \ell'} {\rm p}(\ell_1,\ell_1',\dots, \ell_N,\ell_N') 
    \left| \prod_j g_j(\bm \ell_j,\bm \ell_j ')\right|  \\
    &= \sum_{\bm \ell,\bm \ell'} \frac{1}{\prod_i d_i^2 } 
    \left(\prod_j \prod_{i|E_{ij}=1} d_i \sqrt{  \lambda^{(i)}_{\ell_i} \lambda^{(i)}_{\ell_i'} } \right)\left| \prod_j \bra{\bm \ell_j}_{\bm Q_j} M_{\bm Q_j}^1 \ket{\bm \ell_j'}_{\bm Q_j}\right|.
\end{align}
Recall that in our network there are  two parties connected to each source, hence $\prod_j \prod_{i|E_{ij}=1} d_i \sqrt{  \lambda^{(i)}_{\ell_i} \lambda^{(i)}_{\ell_i'} } =\prod_i d_i^2 \lambda^{(i)}_{\ell_i} \lambda^{(i)}_{\ell_i'} $
and 
\begin{align}
    \E\left [ \left| \prod_j g_j\right| \right]
    &= \sum_{\bm \ell,\bm \ell'} 
   \prod_{i}   \lambda^{(i)}_{\ell_i} \lambda^{(i)}_{\ell_i'} \left| \prod_j \bra{\bm \ell_j}_{\bm Q_j} M_{\bm Q_j}^1 \ket{\bm \ell_j'}_{\bm Q_j}\right| \\
   & = \left|\sum_{\bm \ell ,\bm \ell'}  
   \prod_{i}   \lambda^{(i)}_{\ell_i} \lambda^{(i)}_{\ell_i'} \bra{\bm \ell}  \left(\bigotimes_{j=1}^M M_{\bm Q_j}^1\right)  \ket{\bm \ell'}\right|=\P[a_1=\dots =a_M =1].
\end{align}\\

Now, by the classical Finner inequality, we know that our LV model must satisfy
\be
\P[a_1=\dots =a_M =1] =\E\left [ \left| \prod_j g_j\right| \right] \leq \sqrt{\prod_j \E\left[\left|g_j\right|^2\right]}.
\ee
Hemce, to prove the quantum Finner ineqaulity it remains to show that $\E\left[\left|g_j\right|^2\right]\leq \P[a_j=1]$. Let us show that this is indeed the case. To do so compute
\begin{align}
    \E\left[\left|g_j\right|^2\right]  &= \sum_{\bm \ell,\bm \ell'} \frac{1}{\prod_i d_i^2 } g_j(\bm \ell_j,\bm \ell_j')^2 \\
    &= \sum_{\bm \ell_j,\bm \ell'_j} \frac{1}{\prod_{i|E_{ij}=1} d_i^2 } \left(\prod_{i|E_{ij}=1} d_i \sqrt{  \lambda^{(i)}_{\ell_i} \lambda^{(i)}_{\ell_i'} } \right)^2 \left|\bra{\bm \ell_j}_{\bm Q_j} M_{\bm Q_j}^1 \ket{\bm \ell_j'}_{\bm Q_j}\right|^2 \\
    & = \sum_{\bm \ell_j,\bm \ell'_j} \prod_{i|E_{ij}=1}  \lambda^{(i)}_{\ell_i} \lambda^{(i)}_{\ell_i'}  \bra{\bm \ell_j}_{\bm Q_j} M_{\bm Q_j}^1 \ketbra{\bm \ell_j'}_{\bm Q_j} M_{\bm Q_j}^1  \ket{\bm \ell_j}_{\bm Q_j} \\
    &=\tr \left( \sum_{\bm \ell_j}(\prod_{i|E_{ij}=1}  \lambda^{(i)}_{\ell_i}) \ketbra{\bm \ell_j}_{\bm Q_j}\right) M_{\bm Q_j}^1 \left( \sum_{\bm \ell_j'}(\prod_{i|E_{ij}=1}  \lambda^{(i)}_{\ell_i'}) \ketbra{\bm \ell_j'}_{\bm Q_j}\right) M_{\bm Q_j}^1
\end{align}
Recall that $\sum_{\bm \ell_j}(\prod_{i|E_{ij}=1}  \lambda^{(i)}_{\ell_i})^2 \ketbra{\bm \ell_j}_{\bm Q_j} =\bigotimes_{i|E_{ij}=1} \left(\sum_{\ell_i=1}^{d_i} (\lambda_{\ell_i}^{(i)})^2\ketbra{\ell_i}_{Q_{j}^{(i)}} \right)=\rho_{\bm Q_j}$ is the marginal state recieved by  $A_j$, and since all the state $\ket{\bm \ell_j}$ are orthogonal we also have $\sum_{\bm \ell_j}(\prod_{i|E_{ij}=1}  \lambda^{(i)}_{\ell_i}) \ketbra{\bm \ell_j}_{\bm Q_j} =\sqrt{\rho_{\bm Q_j}}$. This allows us to rewrite the above expression as

\begin{align}
    \E\left[\left|g_j\right|^2\right]  
    &= |\tr \sqrt{\rho_{\bm Q_j}}\,  M_{\bm Q_j}^1 \sqrt{\rho_{\bm Q_j}}\, M_{\bm Q_j}^1| 
    \leq \sqrt{(\tr \sqrt{\rho_{\bm Q_j}}\,  M_{\bm Q_j}^1 M_{\bm Q_j}^1 \sqrt{\rho_{\bm Q_j}}) (\tr M_{\bm Q_j}^1\sqrt{\rho_{\bm Q_j} }\sqrt{\rho_{\bm Q_j}}\,  M_{\bm Q_j}^1)}
    \\
    &=\tr \rho_{\bm Q_j} (M_{\bm Q_j}^1)^2 \leq  \tr \rho_{\bm Q_j} M_{\bm Q_j}^1 
    = \P[a_j=1]
\end{align}
where we used the Cauchy-Schwartz inequality for the Hilbert-Schmidt inner product, and the inequality $(M_{\bm Q_j}^1)^2\leq M_{\bm Q_j}^1$ valid for POVM elements whose eigenvalues are real and between 0 and 1.

\subsection{Quantum Finner rigidity}

\label{app: final proof}

The proof of the quantum Finner inequality can thus by the chain of inequalities

\begin{align}
\P[a_1=\dots =a_M =1] &=\E\left [ \left| \prod_j g_j\right| \right] \\
\label{eq: finner c}
&\leq \sqrt{\prod_j \E\left[\left|g_j\right|^2\right]} = \sqrt{\prod_j \tr \sqrt{\rho_{\bm Q_j}}\,  M_{\bm Q_j}^1 \sqrt{\rho_{\bm Q_j}}\, M_{\bm Q_j}^1} \\
\label{eq: Cauchy-Schwartz}
& \leq \sqrt{\prod_j \tr \rho_{\bm Q_j}  (M_{\bm Q_j}^1)^2 }
\\
\label{eq: proj}
&\leq  \sqrt{\prod_j \tr \rho_{\bm Q_j}  M_{\bm Q_j}^1 } = \sqrt{\prod_j \P[a_j=1]}.
\end{align}

Importantly for us, in the case of equality $\P[a_1=\dots =a_M =1]=\sqrt{\prod_j \P[a_j=1]}$, both bounds (\ref{eq: finner c},\ref{eq: proj}\ref{eq: Cauchy-Schwartz}) must be saturated. We will now use this observation to prove the Theorem.\\

First, note that the equality condition $\tr \rho_{\bm Q_j}  (M_{\bm Q_j}^1)^2=\tr \rho_{\bm Q_j} M_{\bm Q_j}^1$ in Eq.~\eqref{eq: proj} guarantees that all $M_{\bm Q_j}^1=\Pi_{\bm Q_j}^1$ are projectors \footnote{Note that we are only interested and can only reason about the restriction of the measurement operators to the the Hilbert space where the states are supported.}.\\

Second, let us discuss the equality condition of the Cauchy-Schwartz bound in Eq.~\eqref{eq: Cauchy-Schwartz}, for all systems $\bm Q_j$ it must satisfies
\be\label{eq: CS sat}
\tr \sqrt{\rho_{\bm Q_j}}\,  \Pi_{\bm Q_j}^1 \sqrt{\rho_{\bm Q_j}}\, \Pi_{\bm Q_j}^1 = \tr \rho_{\bm Q_j} \Pi_{\bm Q_j}^1.
\ee
This guarantees that $\rho_{\bm Q_j}$ and $\Pi_{\bm Q_j}$ commute.
To see it note that in the rhs one can write $\rho_{\bm Q_j} = \sqrt{\rho_{\bm Q_j}}(\Pi_{\bm Q_j}^1+\id -\Pi_{\bm Q_j}^1)\sqrt{\rho_{\bm Q_j}}$, Eq.~\eqref{eq: CS sat} then implies  $\tr \sqrt{\rho_{\bm Q_j}}\,  \Pi_{\bm Q_j}^1 \sqrt{\rho_{\bm Q_j}}\, (\id -\Pi_{\bm Q_j}^1) = 0
$
showing that $\rho_{\bm Q_j}$ is block-diagonal with respect to the subspace partition $\Pi^1_{\bm Q_1},\id-\Pi^1_{\bm Q_1}$. 
In other words, $\Pi^1_{\bm Q_1}$ must be diagonal in the product basis $\{\ket{\bm \ell_j}\}$, made of the Schmidt bases (of systems $Q_j^{(i)}$) for all the sources connected to $A_j$. Hence they can be written as
\be\label{app:eq:pi_q}
\Pi_{\bm Q_j}^1 = \sum_{\bm \ell_j} \chi_j(\bm \ell_j) \ketbra{\bm \ell_j}_{\bm  Q_j}.
\ee
where $\chi_j(\bm \ell_j) \in \{ 0,1\}$. This implies that the real functions $g_j$ defined in Eq.~\eqref{eq: g def} are of the form
\begin{equation}
    g_j(\bm \ell_j, {\bm \ell}_j') =  \left(\prod_{i|E_{ij}=1} d_i \sqrt{  \lambda^{(i)}_{\ell_i} \lambda^{(i)}_{\ell_i'} } \right) \left|  \bra{\bm \ell_j}_{\bm Q_j} \Pi_{\bm Q_j}^1 \ket{\bm \ell_j'}_{\bm Q_j}\right| = \left(\prod_{i|E_{ij}=1} d_i   \lambda^{(i)}_{\ell_i} \right) \chi_j(\bm \ell_j) \, \delta_{\bm \ell_j, {\bm \ell}_j'}.
\end{equation}

Finally, by the equality condition of the classical Finner inequality in Eq.~\eqref{eq: finner c}, following the Remark \ref{remark:finner_eq_prob}, we also know that the functions $g_j$ must be of the form  
\begin{equation}
g_j(\bm \ell_j, {\bm \ell}_j') = C_j \prod_{i|E_{ij}=1} h_i(\ell_i,\ell'_i)
\end{equation}
with a nonzero constant $C_j$ and non-negative real functions $h_i$ (independent of $j$). Combining the last two equations we find 
\be
 \chi_j(\bm \ell_j) = \bra{\bm \ell_j}_{\bm Q_j} \Pi_{\bm Q_j}^1 \ket{\bm \ell_j}_{\bm Q_j} = \frac{g_j(\bm \ell_j, {\bm \ell}_j)}{\prod_{i|E_{ij}=1} d_i \lambda^{(i)}_{\ell_i}} =C_j \prod_{i|E_{ij}=1} \frac{h_i(\ell_i,\ell_i)}{ d_i \ \lambda^{(i)}_{\ell_i} }=C_j \prod_{i|E_{ij}=1} \tilde h_i(\ell_i)
\ee
Here, since $\chi_j(\bm \ell_j) \in \{ 0,1\}$, we must also have 
$$
\chi_j(\bm \ell_j)  = \prod_{i|E_{ij}=1} \chi^{(i)}(\ell_i) \qquad \text{with } \quad \chi^{(i)}(\ell_i) :=\begin{cases}0 & \tilde h_i(\ell_i) = 0\\1 & \tilde h_i(\ell_i) > 0\end{cases}, $$
and the projector must be product $\Pi_{\bm Q_j}^1  = \bigotimes_{i|E_{ij}=1} \Pi^1_{Q_j^{(i)}}
$, with
\be \label{app:eq:pi_qij}
\Pi^1_{Q_j^{(i)}} = \sum_{\ell_i}   \chi^{(i)} (\ell_i) \ketbra{\ell_i}{\ell_i}
\ee
Note also that in eq \eqref{app:eq:pi_qij}, $\chi^{(i)}$ does not depend on $j$ and for each source $i$ connected to the parties $j$ and $j'$ we have $
\Pi^1_{Q_j^{(i)}}=\Pi^1_{Q_{j'}^{(i)}}$, meaning that they are the same projectors acting on the subsystems  $Q_j^{(i)}$ and $Q_{j'}^{(i)}$.

Finally, note that the model we have just characterized has nothing genuinely quantum to it. Since all the measurements are diagonal in the Schmidt bases of the sources, it is equivalent to a LV model where each source sends an integer $\ell_i$ sampled from $\rm p(\ell_i) =(\lambda_{\ell_i}^{(i)})^2$ to the neighboring parties. And each party $A_j$ outputs $a_j=1$ iff
\be
 \prod_{i|E_{ij}=1} \chi_j^{(i)}(\ell_i)= 1.
\ee

To complete the proof of Theorem \ref{thrm: Q rigidity finner} it remains to show the identity~\eqref{eq: pi}, which follows from straightforward calculation. For the parties $A_{j}$ and $A_{j'}$ connected to a source $S_i$ we find for the probabilities respective probabilities to observe outcomes $1$
\begin{align}
\P[a_j=1] &= \tr \rho_{\bm Q_j} \Pi_{\bm Q_j}^1 = \prod_{i'|E_{i'j}=1} \left(\sum_{\ell_{i'}=1}^{d_{i'}} \chi^{(i')}(\ell_{i'}) (\lambda_{\ell_{i'}}^{(i)})^2\right) = \prod_{i'|E_{i'j}=1} p^{(i')}\\
\P[a_{j'}=1] &= \tr \rho_{\bm Q_{j'}} \Pi_{\bm Q_{j'}}^1 = \prod_{i'|E_{i'j'}=1} \left(\sum_{\ell_{i'}=1}^{d_{i'}} \chi^{(i')}(\ell_{i'}) (\lambda_{\ell_{i'}}^{(i)})^2\right) = \prod_{i'|E_{i'j'}=1} p^{(i')},
\end{align}
where we introduced $p^{(i')} = \sum_{\ell_{i'}=1}^{d_{i'}} \chi^{(i')}(\ell_{i'}) (\lambda_{\ell_{i'}}^{(i)})^2$, which is the marginal probability for the event $\Pi_{Q_j^{i'}}^1$ for both quantum systems controlled by the source $S_i'$. For these two systems these events are however perfectly correlates, and for the joint probability of $a_j=a_{j'}=1$ we find
\begin{align}
\P[a_j=a_{j'}=1] &= \tr (\ketbra{\Psi^{(i)}}_{Q_{j}^{(i)}Q_{j'}^{(i)}} \bigotimes_{i'\neq i|E_{i'j}=1} \rho_{Q_{j}^{(i')}}  \bigotimes_{i''\neq i|E_{i''j'}=1} \rho_{Q_{j'}^{(i'')}}) \Pi_{\bm Q_{j}}^1 \otimes \Pi_{\bm Q_{j'}}^1  \\
& = \prod_{i'|E_{i'j}=1\text{ or }E_{i'j'}=1 } p^{(i')} .
\end{align}
Finally for the ratio we obtain the desired expression
\be
\frac{\P[a_{j}=1] \P[a_{j'}=1] }{\P[a_j=a_{j'}=1]} =\frac{(\prod_{i'|E_{i'j}=1} p^{(i')})(\prod_{i'|E_{i'j'}=1} p^{(i')})}{\prod_{i'|E_{i'j}=1\text{ or }E_{i'j'}=1 } p^{(i')}}=\frac{(p^{(i)})^2}{p^{(i)}}=p^{(i)}=1-e^{(i)}.
\ee
This concludes the proof.

\subsection{Rephrasing the theorem}

\label{app: corrolary}

We have now given the formulation of the theorem in terms of the Schmidt decomposition of purified states prepared by the sources. Which is different from the one given in the main text. Let us discuss how the two are connected.

In the state $\ket{\Psi^{(i)}}_{ Q^{(i)}_j Q^{(i)}_{j'}} =\sum_{\ell_i=1}^{d_i} \lambda_{\ell_i}^{(i)}\ket{\ell_i,\ell_i}_{ Q^{(i)}_j Q^{(i)}_{j'}} $
each "quantum number" $\ell_i$ can be attached a value $\chi_{(i)}(\ell_i)=0$ or $1$ depending on weather $  \Pi_{Q_{j}^{i}}^1 \ket{\ell_i}=0$ or $  \Pi_{Q_{j}^{i}}^1 \ket{\ell_i}=\ket{\ell_i}$. This defines the partition of the Hilbert spaces of the systems ${Q_{j}^{(i)}}$ and ${Q_{j'}^{(i)}}$
in two subspaces defined by the projector $\Pi_{Q_{j}^{(i)}}^1=\Pi_{Q_{j'}^{(i)}}^1$. Hence, the Hilbert space can be decomposed as a tensor product of a qubit, carrying the label $\chi_{(i)}(\ell_i)$, and the remaining degrees of freedom, on which $\Pi_{Q_{j}^{i}}$ acts trivially. For the systems this defines a decomposition $Q_{j}^{(i)} = {\rm Q}_{j}^{(i)} \rm{J}_{j}^{(i)}$. In such a parturition, the state and the projector read 
\begin{align}
\ket{\Psi^{(i)}}_{ Q^{(i)}_j Q^{(i)}_{j'}} &= \sqrt{p^{(i)}} \ket{11}_{{\rm Q}^{(i)}_j {\rm Q}^{(i)}_{j'}} \ket{j^{11}}_{{\rm J}^{(i)}_j {\rm J}^{(i)}_{j'}}+ \sqrt{1-p^{(i)}} \ket{00}_{{\rm Q}^{(i)}_j {\rm Q}^{(i)}_{j'}} \ket{j^{00}}_{{\rm J}^{(i)}_j {\rm J}^{(i)}_{j'}}\\
\Pi_{Q_j^{(i)}}^1&= \ketbra{1}_{\rm{Q}_j^{(i)}} \otimes \id_{\rm{J}_j^{(i)}}
\qquad \Pi_{Q_j^{(i)}}^{\lnot 1}= \ketbra{0}_{\rm{Q}_j^{(i)}} \otimes \id_{\rm{J}_j^{(i)}}
\end{align}
Where we used the fact that we know  the probability $p_i=\bra{\Psi^{(i)}} \Pi^{1}_{Q_j^{(i)}} \ket{\Psi^{(i)}}$, given in Eq.~\eqref{eq: pi}.

Recall that $\ket{\Psi^{(i)}}_{ Q^{(i)}_j Q^{(i)}_{j'}}$ is the purified state, and we have absorbed the purifying system $E_i$ in one of the systems $Q^{(i)}_j Q^{(i)}_{j'}$. 
It remains to see where does it belong with respect to the new partition. This is however straightforward. We know that the measurements performed by the parties must act trivially on the purifying systems, therefore $E^{(i)}$ must be found in the "junk" systems ${\rm J}^{(i)}_j {\rm J}^{(i)}_{j'}$. We can thus rewrite the expression of the state with an explicit mention of the purifying system
\be
\ket{\Psi^{(i)}}_{ Q^{(i)}_j Q^{(i)}_{j'}E^{(i)}} = \sqrt{p^{(i)}} \ket{11}_{{\rm Q}^{(i)}_j {\rm Q}^{(i)}_{j'}} \ket{j^{11}}_{{\rm J}^{(i)}_j {\rm J}^{(i)}_{j'}E^{(i)}}+ \sqrt{1-p^{(i)}} \ket{00}_{{\rm Q}^{(i)}_j {\rm Q}^{(i)}_{j'}} \ket{j^{00}}_{{\rm J}^{(i)}_j {\rm J}^{(i)}_{j'}E^{(i)}},
\ee
and obtain the formulation of the theorem given in the main text.

\section{Photonic entanglement with an SPDC source.}

\label{app: SPDC}

\subsection{Computing the output distribution}

Here we present the detailed analysis of the radomness rate obtained by performing the CHSH test on a photonic entangled state.
More precisely we consider a single-mode polarization entangled SPDC source which prepares a state of the form
\begin{align}
\ket{\Psi} &= \sqrt{\mathcal{N}} \exp(T_1 a_h^\dag b_h^\dag + T_2 a_v^\dag b_v^\dag) \ket{\bm 0} \label{eq app: SPDC state gen}= \sqrt{\mathcal{N}} \exp( \binom{a_h^\dag}{a_v^\dag}^T
{\rm T}\,
\binom{b_h^\dag}{b_v^\dag})\ket{\bm 0}
\end{align}
where $a_{h(v)}$ and $b_{h(v)}$ are the photonic modes sent to Alice and Bob respectively, $\ket{\bm 0}$ is the vacuum state of all the involved modes, and $T_1$ and $T_2$ are parameters controlled by the intensity of the pump laser, with ${\rm T}=
\left(\begin{array}{cc}T_1 & \\ & T_2\end{array}\right)$
and $\mathcal{N} = (1+T_1^2)(1+T_2^2) =\det(\id-{\rm T}^\dag {\rm T})$. One notes that the state on the right hand side in Eq.~\eqref{eq app: SPDC state gen} is well normalize for any complex matrix $\rm T$, such that ${\rm T}^\dag {\rm T}< \id$, which can be seen by performing a singular value decomposition of the matrix and absorbing the rotations in a basis change for the operators $a_h^\dag, a_v^\dag$ and $b_h^\dag, b_v^\dag$. That is we have the following identity
\be
\left\| \exp( \binom{a_h^\dag}{a_v^\dag}^T
T\,
\binom{b_h^\dag}{b_v^\dag})\ket{\bm 0} \right\|^2 = \frac{1}{\det(\id -{\rm M}^\dag {\rm M})} \qquad \forall \,{\rm M}:\,  {\rm M}^\dag {\rm M}<\id
\ee

The state $\ket{\Psi}$ produced by the source is distributed to the two parties Alice and Bob, who can apply local linear-optical transformations of the polarization of the their respective modes, described by the following linear transformations of the creation operators, parametrized by the angles $\alpha,\varphi_A$ and $\beta,\varphi_B$ respectively 
\begin{align}
    U_A \binom{a_h^\dag}{a_v^\dag} U_A^\dag &= R(\alpha,\varphi_A,\phi_A)\binom{a_h^\dag}{a_v^\dag}  \qquad \text{with} \qquad R(\theta,\varphi,\phi ) :=  \left(\begin{array}{cc} 
\cos(\theta) e^{\ii \phi} &  \sin(\theta) e^{\ii \varphi}\\
-e^{-\ii \varphi} \sin(\theta)&  \cos(\theta) e^{-\ii \phi}
\end{array}\right)\\
 U_B \binom{b_h^\dag}{b_v^\dag} U_B^\dag &= R(\beta,\varphi_B,\phi_B)\binom{b_h^\dag}{b_v^\dag} ,
 \end{align}
 where $R(\theta,\varphi,\phi )$ is a general element of the special unitary group of degree 2.  The polarization rotation leads to the following state transformation
\begin{align}
\ket{\tilde\Psi_{\alpha,\varphi_A,\beta,\varphi_B}} &=
U_A\otimes U_B \ket{\Psi} \\
&= 
 \sqrt{\mathcal{N}} \exp(U_A\otimes U_B ( T_1 a_h^\dag b_h^\dag + T_2 a_v^\dag b_v^\dag) U_A^\dag\otimes U_B^\dag) \ket{\bm 0} \\
 & =\sqrt{\mathcal{N}} \exp( \binom{a_h^\dag}{a_v^\dag}^T \underbrace{R^T(\alpha,\varphi_A, \phi_B)
\left(\begin{array}{cc} 
T_1 & \\
& T_2 
\end{array}\right) R(\beta,\varphi_B,\phi_B)}_{:= {\rm M}(T_1,T_2,\alpha,\varphi_A,\phi_A,\beta,\varphi_B,\phi_B)}
\binom{b_h^\dag}{b_v^\dag})\ket{\bm 0}
\end{align}
For example, the state is particularly simple for $T=T_1=T_2$ and $\varphi_A=\phi_A=\varphi_B=\phi_B=0$ where 
\begin{align}
{\rm M}(T,T,\alpha,0,\beta,0)=T \left(
  \begin{array}{cc}
 \cos (\alpha -\beta ) & - \sin (\alpha -\beta ) \\
  \sin (\alpha -\beta ) &  \cos (\alpha -\beta ) \\
\end{array}
\right)
\end{align}

The number of photons in each mode $a_h,a_v, b_h, b_v$ are then measured as we will now describe. But first, it is important to mention that such measurement are insensitive to phase, i.e. any transformation of the form $a^\dag \mapsto a^\dag e^{\ii \xi}$. Therefore, the two polarization transformation
\begin{align}
    R(\alpha,\varphi,\phi)\binom{a_h^\dag}{a_v^\dag}  &= \binom{\cos(\alpha) e^{\ii \phi} a_h^\dag +\sin(\theta) e^{\ii \varphi} a_v^\dag}{- \sin(\theta)e^{-\ii \varphi} a_h^\dag + \cos(\alpha) e^{-\ii \phi} a_v^\dag} \qquad \text{and}\\
    R(\alpha,\varphi-\phi,0)\binom{a_h^\dag}{a_v^\dag} &= \binom{\cos(\alpha) a_h^\dag +\sin(\theta) e^{\ii (\varphi-\phi)} a_v^\dag}{- \sin(\theta)e^{-\ii (\varphi-\phi)} a_h^\dag + \cos(\alpha)  a_v^\dag}
\end{align}
are related by such phase transformation give rise to the same measurement statistics. Without loss of generality when optimizing the parameters we can thus set $\phi_A=\phi_B=0$ and write $R(\theta,\varphi,0)=R(\theta,\varphi)$, ${\rm M}(T_1,T_2,\alpha,\varphi_A,\phi_A,\beta,\varphi_B,\phi_B)={\rm M}(T_1,T_2,\alpha,\varphi_A,\beta,\varphi_B)$ and $\ket{\tilde\Psi}=\ket{\tilde\Psi_{\alpha,\varphi_A,\beta,\varphi_B}}$.

On each side we can detect no photons, one photon or both photons, there four events define the following POVM
\begin{align}
E_{\circ\circ}^{A} &= D_{\circ}^{(a_h)}\otimes D_{\circ}^{(a_v)} = \ketbra{00}_{hv}\\
E_{\bullet\circ }^{A} &=  D_{\bullet}^{(a_h)}\otimes D_{\circ}^{(a_v)}=(\id-\ketbra{0})_h \otimes \ketbra{0}_v \\
E_{ \circ\bullet}^{A}  &=   D_{\circ}^{(a_h)}\otimes D_{\bullet}^{(a_v)}=\ketbra{0}_h \otimes (\id-\ketbra{0})_v \\ 
E_{\bullet \bullet}^{A}  & =   D_{\bullet}^{(a_h)}\otimes D_{\bullet}^{(a_v)}=(\id-\ketbra{0})_h \otimes (\id-\ketbra{0})_v,
\end{align}
and the same for $E^B_b$ at Bob. The output probability distribution is thus given by
\be
{\rm P}(a,b|\alpha,\varphi_A,\beta, \varphi_B) = \bra{\tilde\Psi_{\alpha,\varphi_A,\beta,\varphi_B}} E_a^A\otimes E_b^B \ket{\tilde\Psi_{\alpha,\varphi_A,\beta,\varphi_B}}
\ee
where $a,b\in \{\circ\circ,\bullet \circ, \circ \bullet,\bullet\bullet\}$ and the state has an implicit dependence on $T_1$ and $T_2$. Let us denote each outcome with $a = h_A v_A$ with $h_A= 1$ if the corresponding detector clicks ($\bullet$) and $h_A=0$ if it does not ($\circ$), and rewrite the output probabilities
\be
 {\rm P}(h_A,v_A,h_B,v_B|\alpha,\varphi_A,\beta, \varphi_B) = \bra{\tilde\Psi_{\alpha,\varphi_A,\beta,\varphi_B}} D_{h_A}^{(a_h)} \otimes D_{v_A}^{(a_v)} \otimes D_{h_B}^{(b_h)} \otimes D_{v_B}^{(b_h)} \ket{\tilde\Psi_{\alpha,\varphi_A,\beta,\varphi_B}}
\ee
where $D_{h_A} =\begin{cases}\ketbra{0} & h_A=0\\
\id - \ketbra{0} & h_A =1 \end{cases}$. From this expression it is clear that by expanding the terms $(\id - \ketbra{0})$, all the probabilities can be computed as a linear combination of the following terms 
\begin{align}
 \left\langle G_{x}^{(a_h)} \otimes G_{y}^{(a_v)} \otimes G_{z}^{(b_h)} \otimes G_{w}^{(b_h)} \right\rangle &:= \bra{\tilde\Psi_{\alpha,\varphi_A,\beta,\varphi_B}} G_{x}^{(a_h)} \otimes G_{y}^{(a_v)} \otimes G_{z}^{(b_h)} \otimes G_{w}^{(b_h)} \ket{\tilde\Psi_{\alpha,\varphi_A,\beta,\varphi_B}}   \\
 & =\left \|G_{x}^{(a_h)} \otimes G_{y}^{(a_v)} \otimes G_{z}^{(b_h)} \otimes G_{w}^{(b_h)} \ket{\tilde\Psi_{\alpha,\varphi_A,\beta,\varphi_B}} \right \|^2
\end{align}
 where $G_x =\begin{cases}\ketbra{0} & x=0\\
\id & x =1 \end{cases}$ and we used $G_x^2 =G_x$. These terms can be computed relatively easily as we find 
\begin{align}
    &G_{x}^{(a_h)} \otimes G_{y}^{(a_v)} \otimes G_{z}^{(b_h)} \otimes G_{w}^{(b_h)} \ket{\tilde\Psi_{\alpha,\varphi_A,\beta,\varphi_B}} \\
    &= \sqrt{\mathcal{N}}   G_{x}^{(a_h)} \otimes G_{y}^{(a_v)} \otimes G_{z}^{(b_h)} \otimes G_{w}^{(b_h)} \exp( \binom{a_h^\dag}{a_v^\dag}^T  {\rm M}(T_1,T_2,\alpha,\varphi_A,\beta,\varphi_B)
\binom{b_h^\dag}{b_v^\dag})\ket{\bm 0}\\
& =\sqrt{\mathcal{N}}  \exp( \binom{ a_h^\dag}{ a_v^\dag}^T 
\left(\begin{array}{cc}
x & \\
& y 
\end{array}\right)
{\rm M}(T_1,T_2,\alpha,\varphi_A,\beta,\varphi_B) \left(\begin{array}{cc}
z & \\
& w
\end{array}\right)
\binom{b_h^\dag}{b_v^\dag})\ket{\bm 0},
\end{align}
leading to 
\begin{align}
 \left\langle G_{x}^{(a_h)} \otimes G_{y}^{(a_v)} \otimes G_{z}^{(b_h)} \otimes G_{w}^{(b_h)} \right\rangle &=  \frac{\det(\id-{\rm T}^\dag {\rm T})}{\det(\id -{\rm M}^\dag \left(\begin{array}{cc}
x & \\
& y 
\end{array}\right){\rm M}\left(\begin{array}{cc}
z & \\
& w 
\end{array}\right))},
\end{align}
which is straightforward to compute. We have this derived an analytic expression for ${\rm P}(a,b|\alpha,\varphi_A,\beta, \varphi_B)$ in terms of the parameters $\alpha,\varphi_A,\beta, \varphi_B, T_1$ and $T_2$. 

\subsection{Applying the Finner rigidity}

As discussed in the main text we now identify the outcomes $a,b=\circ \circ$ to be the failure outcomes $\varnothing$. Since the detectors are ideal, it is clear that these outcomes are perfectly correlated and only occur when the quantum source emits the vacuum state $\ket{\bm 0}$.
Hence we find that the Finner inequality is saturated 
\be
\P[a,b\neq\varnothing] =\P[a\neq\varnothing]= \P[b\neq\varnothing] = \sqrt{\P[a\neq\varnothing] \P[b\neq\varnothing] },
\ee
and by Corollary \ref{corrolary} post-selected does not open the detection loophole. More precisely, the quantum model underlying the post-selected distribution (including the purifying system behold by an eavesdropper)
$$
\widetilde{\rm P}(a,b|\alpha,\varphi_A,\beta, \varphi_B) =1-\frac{{\rm P}(a,b,a\neq \varnothing,b\neq \varnothing |\alpha,\varphi_A,\beta, \varphi_B)}{\P[a,b\neq\varnothing]}
$$
must be consisted with the network structure.

For the probability of conclusive outcomes we find 
\be
\P[a,b\neq\varnothing]= 1 -\P[a=b=\varnothing] = 1- \left\langle G_{0}^{(a_h)} \otimes G_{0}^{(a_v)} \otimes G_{0}^{(b_h)} \otimes G_{0}^{(b_h)} \right\rangle =
1- \det(\id-{\rm T}^\dag {\rm T}) = 1- (1-T_1^2)(1-T_2^2).
\ee

\subsection{Optimizing the CHSH violation}

The CHSH test 
\be
{\rm CHSH}[{\rm P}(a,b|x,y)]= \sum_{a,b,x,y=0,1} (-1)^{a+b +xy} \,\, {\rm P}(a,b|x,y)
\ee
involve two binary measuremnts per party. Hence to compute the CHSH score in our scenario we need to select two pairs of measurement settings $(\alpha,\varphi_A)$ with $(\alpha',\varphi_A)$ and $(\beta,\varphi_B)$ with $(\beta',\varphi_B')$, and furthermore post-process the outputs $a$ with $b$ from quaternary (ternary with post-selection) to binary values, by binning some outputs together. Furthermore, we obviously wish to maximize the CHSH score with respect to the settings choice, the binning strategy (discrete) and the source parameters $T_1$ and $T_2$. 

For the bare distribution $\rm P$ computed in the previous sections we can thus define the optimal CHSH to be
\begin{equation}
    {\rm CHSH}^* = \max_{\rm binning}\quad  \max_{T_1,T_2 \in [0,1)} \quad \max_{\alpha,\varphi_A,\alpha',\varphi_A',\beta,\varphi_B,\beta',\varphi_B'} {\rm CHSH}[\rm P].
\end{equation}
Performing the optimization numerically we find $ {\rm CHSH}^* =2.3057 $ given in Table~\ref{tab:optimization_results}(left) together with the parameter values. As one would expect for the post-selection distribution $\widetilde{\rm P}$ the CHSH score can be made arbitrarily close to the quantum bound of $2\sqrt{2}$ as the price of decreasing post-selection probability $\P[a,b\neq\varnothing]$ (denoted $\rm P\_succ$ in the table). This optimization is thus not very useful.

To further compare the two regimes in more details we also considered the specific case where $T_1=T_2=T$. Here we found that the optimal bare CHSH score is slightly lower ${\rm CHSH}^* =2.3008$, see Table~\ref{tab:optimization_results}(right). For fixed $T$ the optimization of the post-selected CHSH score is nontrivial (it only give $2\sqrt{2}$ for $T\to 0$), and in Fig.~\ref{fig:square_plot} we present a comparison of the optimal bare CHSH and post-selected CHSH scores. A short-coming of this comparison is that the CHSH score completely ignores the post-selection rate. For a more "fair" comparison we consider the rate of device-independent randomness generated in the two scenarios. \\

\begin{table}[ht]
    \centering
    \begin{minipage}[t]{0.48\textwidth}
        \centering
        \begin{tabular}{@{}lcccccc@{}}
        \toprule
        \multicolumn{7}{c}{\textbf{Optimization Results for general $T_1$ and $T_2$}} \\ \midrule
        \textbf{Setting}      & {T1}   & {T2}   & {$\alpha_0$} & {$\alpha_1$} & {$\beta_0$} & {$\beta_1$} \\ \midrule
        \textbf{Standard}     & 0.6098 & 0.7148 & 1.6684 & 4.3063 & 4.5638 & 1.9214 \\
        CHSH Score            & \multicolumn{6}{c}{2.3057} \\
        Randomness            & \multicolumn{6}{c}{0.2525} \\ \midrule
        \textbf{Post-select}  & 0.6308 & 0.5799 & 3.2632 & 2.6783 & 2.9730 & 3.5572 \\
        CHSH Score            & \multicolumn{6}{c}{2.5013} \\
        P\_succ               & \multicolumn{6}{c}{0.6004} \\
        Randomness            & \multicolumn{6}{c}{0.2749} \\ \bottomrule
        \end{tabular}
    \end{minipage}
    \hfill
    \begin{minipage}[t]{0.48\textwidth}
        \centering
        \begin{tabular}{@{}lcccc@{}}
        \toprule
        \multicolumn{5}{c}{\textbf{Optimization Results for  $T_1 = T_2 =T$}} \\ \midrule
        \textbf{Setting}      & {T} & {$\alpha_0$} & {$\alpha_1$} & {$\beta_1$} \\ \midrule
        \textbf{Standard}     & 0.6379 & 3.4351 & 2.8472 & 3.7285 \\
        CHSH Score            & \multicolumn{4}{c}{2.3008} \\
        Randomness            & \multicolumn{4}{c}{0.2479} \\ \midrule
        \textbf{Post-select}  & 0.5721 & 2.8240 & 3.4587 & 2.5074 \\
        CHSH Score            & \multicolumn{4}{c}{2.5326} \\
        P\_succ               & \multicolumn{4}{c}{0.5474} \\
        Randomness            & \multicolumn{4}{c}{0.2712} \\ \bottomrule
        \end{tabular}
    \end{minipage}
    \caption{Optimization results for maximizing the randomness rate}
    \label{tab:optimization_results}
\end{table}

\subsection{Bounding the randomness}

The randomness of a given output used in the CHSH test, can be directly connected to the CHSH score. For concreteness let us label ${\rm B}_1$ to be the classical random variable encoding the outcome of Bob's measurement for the setting $y=0$ (i.e. $\beta, \varphi_B$). The device-independent randomness carried by this random variable can by quantified the conditional von Neumann entropy $H({\rm B}_1|E)$, where in general the system $E$ behold by the eavesdropper is the purification of the states prepared by the sources. In the context of device-independent quantum key distribution~\cite{pironio2009device} with collective attacks it has been shown that this quantity is directly related to the CHSH score 
$$
H({\rm B}_1|E) \geq \left(1-h \left( \frac{1 + \sqrt{( {\rm CHSH}/2 )^2 - 1}}{2} \right)\right),
$$
where $h$ is the binary entropy. Note that in~\cite{pironio2009device}, the bound was derived assuming uniform marginals, in particular $H({\rm B}_1)=1$, which can always be achieved by performing a symmetrization step (using local randomness and public comminication, which are free resources in QKD). Nevertheless, it was later observed that the bound is also valid without this step, see Appendix A in \cite{ho2020noisy}.

For the bare distribution $\rm P$ the randomness rate is thus given by $R=\left(1-h \left( \frac{1 + \sqrt{( {\rm CHSH}/2 )^2 - 1}}{2} \right)\right)$. In turn for the post-selected distribution $\widetilde{\rm P}$, the randomness rate must also take into account the fraction $\P[a,b\neq\varnothing]= 1- (1-T_1^2)(1-T_2)^2$ of rounds that are kept after post-selection, so that we have 
$$
\widetilde R = \left(1-h \left( \frac{1 + \sqrt{( \widetilde{\rm CHSH}/2 )^2 - 1}}{2} \right)\right)  \left( 1- (1-T_1^2)(1-T_2^2) \right).
$$

\begin{figure}[H]
    \centering
    \includegraphics[width=0.49\textwidth]{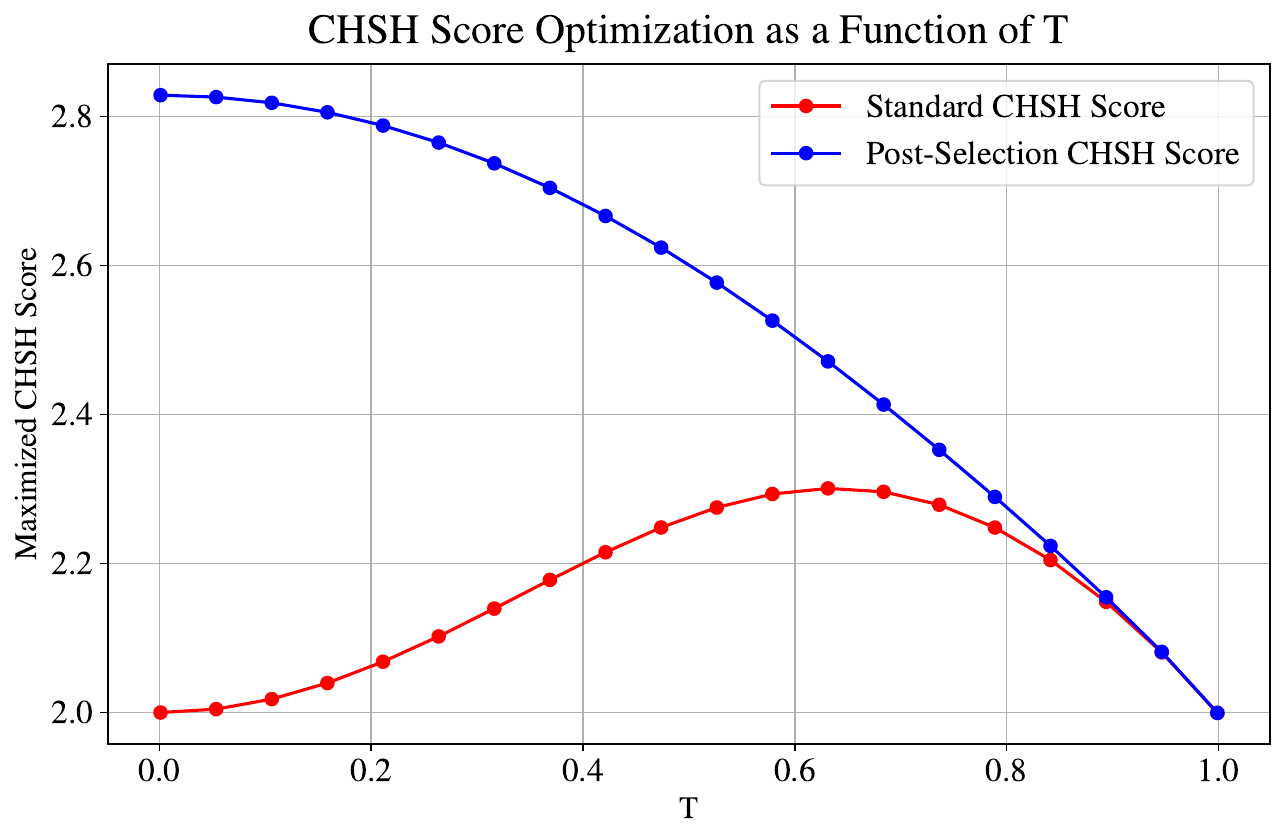}\quad
    \includegraphics[width=0.49\textwidth]{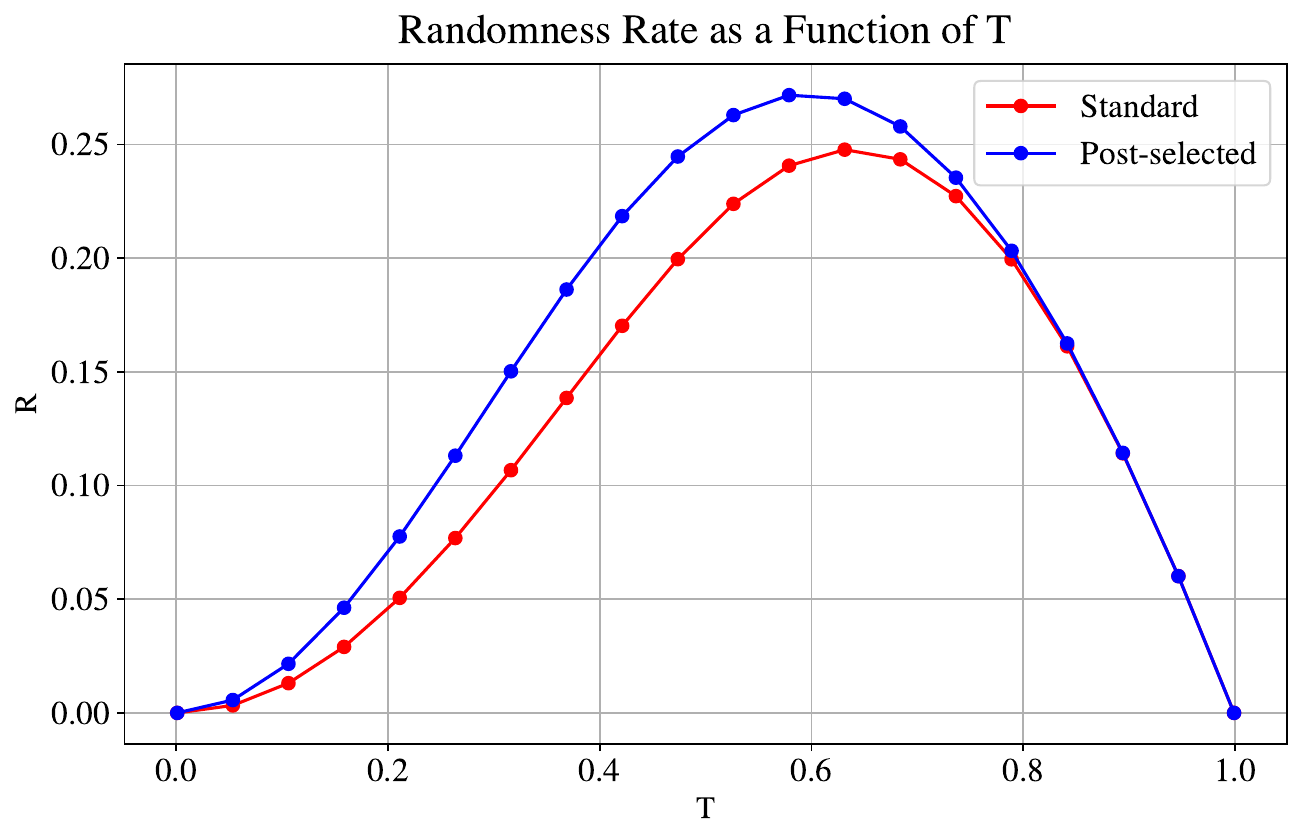}
    \caption{\textbf{Bell Scenario: } Comparison of CHSH score and Randomness plots for Standard and post selected statistics}%
    \label{fig:square_plot}%
\end{figure}

Now both quantities $R$ and $\widetilde R$ can be optimized meaningfully over the free parameters. In Tables I and II we give the results for the global optimization. And in Fig.~\ref{fig:square_plot} we plot the two randomness rates as a function of $T=T_1=T_2$. We see that the post-selected approach achieves a higher rate.

To conclude, we note that the presented results are quire promising regarding the usefulness of the Finner rigidity, however they assume iid attacks, infinite statistics effects, and ideal detectors. At this point we believe that the main thing required to go further is to establish an approximate quantum Finner rigidity bound, which would allow to partially self-test the setup in the case where the Finner inequality is not exactly saturated.

\section{Nonlocality and randomness in the triangle network}\label{app: RGB4}

\begin{figure}[t]
    \centering
    \includegraphics[width=0.3\linewidth]{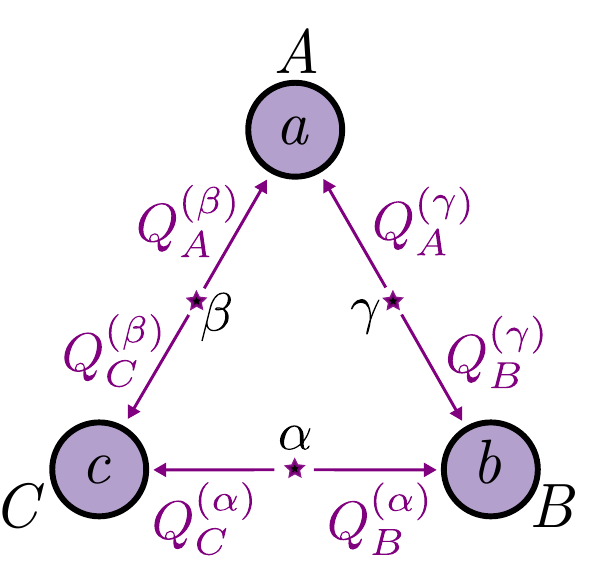}
    \caption{triangle network} 
    \label{fig:triangle}
\end{figure}

Here, we consider the triangle network scenario, consisting of three parties, A, B, and C, pairwise connected by bipartite sources labeled $\alpha, \beta,\gamma$ according to Fig. \ref{fig:triangle}. Each party receives two systems from the neighboring sources and produces an output 
$a, b, c$. Overall, there are six systems involved, labeled as $Q^{(\xi)}_X$ where $\xi \in \{\alpha, \beta,\gamma\}$ denotes the source generating it and $X\in \{ A,B,C\}$ identifies the connected party receiving the system. The global state of the system admits a purification 
$\ket{\Psi}=\ket{\psi^{\alpha}}_{Q^{(\alpha)}_B Q^{(\alpha)}_C E^{(\alpha)}} \ket{\psi^{\beta}}_{Q^{(\beta)}_C Q^{(\beta)}_A E^{(\beta)}}\ket{\psi^{\gamma}}_{Q^{(\gamma)}_A Q^{(\gamma)}_B E^{(\gamma)}}$ 
and the measurements performed by each party are characterized by the POVMs $\{M_A^a\}_a$,$\{M_B^b\}_b$,$\{M_C^c\}_c$  acting on the pairs of systems $Q^{(\beta)}_A Q^{(\gamma)}_A,  Q^{(\gamma)}_B Q^{(\alpha)}_B, Q^{(\alpha)}_C $

Let us assume that each source fails with probability $e^{(\xi)}$, resulting in neighboring parties reporting a failure outcome $\varnothing$. First, we observe that the resulting correlations will saturate the Finner inequality~\eqref{eq:finner_eq}, and by Theorem~\ref{thm: finner st} the states prepared by the sources must be of the following form 

\begin{align}\label{appendix: state rigid}
\ket{\Psi^{(\xi)}}_{Q_X^{(\xi)}Q_{Y}^{(\xi)}E^{(\xi)}} \!\! &= \sqrt{1-e^{(\xi)}} \ket{11}_{{F}^{(\xi)}_X{F}^{(\xi)}_{Y}}\ket{\widetilde{\Psi}^{(\xi)}}_{{ \widetilde Q}^{(\xi)}_X { \widetilde Q}^{(\xi)}_{Y}E^{(\xi)}}\nonumber  \\
&\!\!\!+\sqrt{e^{(\xi)}} \ket{00}_{{F}^{(\xi)}_X{F}^{(\xi)}_{Y}}\ket{\Theta^{(\xi)}}_{{ \widetilde Q}^{(\xi)}_X { \widetilde Q}^{(\xi)}_{Y}E^{(\xi)}}
\end{align}
where, $X, Y$ indicate the  parties connected to the source $\xi$ and each system $Q_{X}^{(\xi)}=F_{X}^{(\xi)}\widetilde Q_{X}^{(\xi)}$  is decomposed in the flag qubit $F_{X}^{(\xi)}$ and the remaining degrees of freedom $\widetilde Q_{X}^{(\xi)}$, $E^{(\xi)}$ is the system purifying the source. Moreover, the measurements are of the form

\begin{align}\label{appendix: meas finner}
\begin{cases} 
M_{X}^{x} = \ketbra{11}_{F_{X}^{(\xi)}F_{X}^{(\xi')}}\otimes \widetilde{M}_{X}^{x} & x \not = \varnothing \\
M_{X}^{\varnothing} =\big (\ketbra{00}+ \ketbra{01}+\ketbra{10}\big)_{F_{X}^{(\xi)}F_{X}^{(\xi')}}  \otimes \id_{\widetilde Q_{X}^{(\xi)}\widetilde Q_{X}^{(\xi')}} & 
\end{cases}.
\end{align}
Here, $\xi, \xi'$ indicate the sources connected to the party $X$, and $\widetilde{M_{X}^{x}}$ is a POVM with only conclusive outcomes acting on 
$\widetilde Q_{X}^{(\xi)}\widetilde Q_{X}^{(\xi')}$. Hence, all the outputs are conclusive iff all the measurements $M_X^x$ project the flag qubits the flag qubits $F_{X}^{(\xi)}F_{X}^{(\xi')}$ onto $\ket{11}$. Following the results of this paper, the post-selected statistics $\widetilde{\rm P}(a,b,c) = \rm P(a,b,c|a,b,c \not = \varnothing)$ is consistent with the triangle network and must be achieved with the states $\ket{\widetilde{\Psi}^{(\xi)}}$ and measurements $\widetilde{M_{X}^{x}}$.\\

Now assume that for $u_0=-v_1=\cos(\theta)$ and $v_0= u_1= \sin(\theta)$ with $\theta \in [0,\pi/4]$, the  distribution $\widetilde{\rm P}(a,b,c)$  is of the form of the RGB4 distribution, with $a,b,c \in \{ 0, 1_0, 1_1 , 2\}$
\be\label{eq: RGB$ def}\begin{split}
&P_{RGB4}( 1_i, 1_j , 1_k) = \frac{1}{8} (u_i u_j u_k + v_i v_j v_k)^2\\
&P_{RGB4}(1_i, 0,2)=\frac{1}{8} u_i^2, \quad
P_{RGB4}(1_i, 2,0)= \frac{1}{8} v_i^2 \quad \circlearrowright
\end{split}
\ee
 where $\circlearrowright$ means that the equation is valid up to cyclic permutations of the parties. Then following the results of ref. \cite{Sekatski2023} the systems $\widetilde Q_X^{(\xi)}$ contain a qubit subsystem $\bar Q_X^{(\xi)}$ such that the states  $\ket{\widetilde{\Psi}^{(\xi)}}_{\widetilde Q_X^{(\xi)}\widetilde Q_{Y}^{(\xi)}E^{(\xi)}}$ and the measurements $\widetilde{M_{X}^{x}}$ are of the form
\begin{align}
 \ket{\widetilde{\Psi}^{(\xi)}}_{\widetilde Q_X^{(\xi)}\widetilde Q_{Y}^{(\xi)}E^{(\xi)}} &= \frac{1}{\sqrt{2}}\big( \ket{01}_{\bar Q_{X}^{(\xi)} \bar Q_{Y}^{(\xi)}}\!\ket{j^{c}_\xi}_{J^{(\xi)}}\!\!+
\ket{10}_{\bar Q_{X}^{(\xi)} \bar  Q_{Y}^{(\xi)}}\ket{j^{a}_\xi}_{J^{(\xi)}}
 \big)
 \\
\widetilde{M_{X}^{0}} &= \ketbra{00}_{\bar Q_{X}^{(\xi)} \widetilde Q_{X}^{(\xi' )}}\otimes \mathds{1}_{J_X} \nonumber \\
\widetilde {M_{X}^{2}}&=\ketbra{11}_{\bar Q_{X}^{(\xi)} \bar Q_{X}^{(\xi')}}\otimes \mathds{1}_{J_X} \\
 \widetilde{M_{X}^{1}}&=\widetilde{M_{X}^{1_0}}+\widetilde{M_{X}^{1_1}}=(\ketbra{01}+\ketbra{10})_{\bar Q_{X}^{(\xi)} \bar Q_{X}^{(\xi')}}\otimes \mathds{1}_{J_X} \nonumber.
\end{align}
where the junk systems $J_X$ contains the reaming degrees of freedom in $\widetilde Q_{X}^{(\xi)}\widetilde Q_{X}^{(\xi')}$, $J^{(\xi)}$ contains the remaining degrees of freedom in $\widetilde Q_{X}^{(\xi)} \widetilde Q_{Y}^{(\xi)}$ as well as the purifying system $E^{(\xi)}$, and $\ket{j^{c/a}_\xi}_{J^{(\xi)}}$ are some unknown states thereof. We now focus on the outcome of a single party, say $A$. After coarse-graining it to three values 
\begin{equation}
\bar a = \begin{cases} 0 &a=0,2 \\
1 & a=1_0,1_1 \\
\varnothing & a = \varnothing
\end{cases}
\end{equation}
let it be recorder in a classical register, denoted $\bar{\rm A}$. The classical-quantum state of this register and the purifying systems $E^{(\beta)}$ and $E^{(\gamma)}$, potentially controlled by an eavesdropper, is given by (for $\rm p_A =(1-e^{(\beta)})(1-e^{(\gamma)})$)
\begin{align}
\varrho_{\bar{\rm A}E^{(\beta)}E^{(\gamma)}} &= \rm p_A \left( \frac{1}{2}  \ketbra{0}_{\bar{\rm A}}\otimes\rho_{E^{(\beta)}E^{(\gamma)}}^{(0)}+\frac{1}{2}  \ketbra{1}_{\bar{\rm A}}\otimes\rho_{E^{(\beta)}E^{(\gamma)}}^{(1)} \right) + (1-\rm p_A ) \ketbra{\varnothing}_{\bar{\rm A}}\otimes\rho_{E^{(\beta)}E^{(\gamma)}}^{(\varnothing)}\quad 
\end{align}
with
\begin{align}
\rho_{E^{(\beta)}E^{(\gamma)}}^{(0)} & = \text{tr}_{\lnot E^{(\beta)}E^{(\gamma)}} \,(\widetilde{M_{A}^{0}}+\widetilde{M_{A}^{2}}) \ketbra{\widetilde{\Psi}^{(\beta)},\widetilde{\Psi}^{(\gamma)}}\\
\rho_{E^{(\beta)}E^{(\gamma)}}^{(1)} & =\text{tr}_{\lnot E^{(\beta)}E^{(\gamma)}} \,(\widetilde{M_{A}^{1_0}}+\widetilde{M_{A}^{1_1}}) \ketbra{\widetilde{\Psi}^{(\beta)},\widetilde{\Psi}^{(\gamma)}}
\end{align}
where $\text{tr}_{\lnot E^{(\beta)}E^{(\gamma)}}$ means that we trace out all the degrees of freedom excepts the purifying systems $E^{(\beta)}E^{(\gamma)}$, and we used Eqs.~\eqref{appendix: state rigid} and \eqref{appendix: meas finner}. The randomness of the output $\bar a$ (from the eavesdropper perspective), is naturally quantified by the conditional von Neumann entropy $H(\bar{\rm A}|E)_{\varrho_{\bar{\rm A}E^{(\beta)}E^{(\gamma)}}}$ of this classical-quantum state. The data precessing inequality guarantees that
\be
H(\bar{\rm A}|E) \geq {\rm p}_A \, H(\bar{\rm A}|E, \bar a \neq \varnothing) + (1-{\rm p}_A)\underbrace{H(\bar{\rm A}|E, \bar a = \varnothing)}_{=0},  
\ee
meaning that the eavesdropper could only increase theirs knowledge of $\bar {\rm A}$ by knowing if the output was conclusive. Here, the first term $H(\bar{\rm A}|E, \bar a \neq \varnothing)_{\varrho_{\bar{\rm A}E^{(\beta)}E^{(\gamma)}}} = H(\bar{\rm A}|E)_{\widetilde\varrho_{\bar{\rm A}E^{(\beta)}E^{(\gamma)}}}$ is simply the conditional entropy of the state  in the conclusive-output branch, i.e. the state
\be
\widetilde \varrho_{\bar{\rm A}E^{(\beta)}E^{(\gamma)}} =  \frac{1}{2}  \ketbra{0}_{\bar{\rm A}}\otimes\rho_{E^{(\beta)}E^{(\gamma)}}^{(0)}+\frac{1}{2}  \ketbra{1}_{\bar{\rm A}}\otimes\rho_{E^{(\beta)}E^{(\gamma)}}^{(1)}
\ee
leading the (post-selected) RGB4 distribution. A bound on this entropy  has been derived in \cite{Sekatski2023} and subsequently improved in  \cite{boreiri2024topologically}. Combining the two results we obtain 
\begin{align}
 H(\bar{\rm A}|E)_{\widetilde\varrho_{\bar{\rm A}E^{(\beta)}E^{(\gamma)}}}&\geq L:= 1 + h\left(\frac{1+4r}{2}\right) - H\left(
        \frac{(1+\sqrt{4r})^2}{4},
        \frac{(1-\sqrt{4r})^2}{4},
        \frac{1-4r}{4},
        \frac{1-4r}{4}\right)
\end{align}

Where $h(x)$ is the entropy of the binary distribution $(x, 1-x)$ and $H$ is the entropy of a four-output distribution and the parameter $r$ quantifies the coherence of the observed distribution and can be lower bounded by \cite{Sekatski2023}
$$r \geq \frac{1}{2} \sin^3(\theta ) \big(3 \cos (\theta )+\cos (3 \theta )-6 \sin (\theta ) \big).$$ 
Coming back to the full distribution we can therefore conclude that 
\begin{equation}
H(\bar{\rm A}|E)_{\varrho_{\bar{\rm A}E^{(\beta)}E^{(\gamma)}}}\geq (1-e^{(\beta)}) (1-e^{(\gamma)}) L 
\end{equation}
\bigskip

Moreover, it was demonstrated in \cite{Sekatski2023} that observing the RGB4 distribution, as defined in Eq. \ref{eq: RGB$ def}, ensures that the measurements of all parties are non-separable and that all sources distribute entanglement among the parties. Both conclusions clearly persist here, despite the presence of unreliable sources. 

\newpage
\twocolumngrid

\bibliographystyle{unsrt}
\bibliography{main.bib}

\end{document}